\documentclass[12pt]{article}
\usepackage{amsfonts}
\usepackage{amssymb}
\usepackage{amsmath}
\usepackage{graphicx}
\usepackage{psfrag}
\usepackage{epsfig}

\renewcommand{\baselinestretch}{1.1}

\def\IR{\mathbb{R}}
\def\IC{\mathbb{C}}
\def\IZ{\mathbb{Z}}

\def\Tr{\mbox{Tr}}

\def\id{\mbox{id}}

\def\be{\begin{equation}}
\def\ee{\end{equation}}
\def\bea{\begin{eqnarray}}
\def\eea{\end{eqnarray}}

\font\mybb=msbm10 at 12pt

\def\bb#1{\hbox{\mybb#1}}

\def\Z {\bb{Z}}

\def\Div{{\rm Div}}

\def\ibar{{\bar\imath}}
\def\jbar{{\bar\jmath}}

\def\S{ section }

\def\id{\protect{{1 \kern-.28em {\rm l}}}}

\makeatletter
\renewcommand\section{\@startsection {section}{1}{\z@}%
                                   {-3.5ex \@plus -1ex \@minus -.2ex}%
                                   {2.3ex \@plus.2ex}%
                                   {\normalfont\large\bfseries}}

\renewcommand\subsection{\@startsection{subsection}{2}{\z@}%
                                   {-3.25ex\@plus -1ex \@minus -.2ex}%
                                   {1.5ex \@plus .2ex}%
       {\normalfont\normalsize\bfseries}}
\makeatother

\begin{document}

\begin{titlepage}

\begin{flushright}
hep-th/0505099\\
PUPT-2160\\
ITEP-TH-33/05\\
\end{flushright}

\vspace{5mm}

\begin{center}
{\huge Perturbative Search for Fixed Lines}
\\
\vspace{3mm}
{\huge in
Large $N$ Gauge Theories
}\\
\vspace{1mm}
\end{center}

\vspace{1mm}
\begin{center}
{\large A. Dymarsky, I.R. Klebanov and R. Roiban
}\\
\vspace{3mm}
Joseph Henry Laboratories, Princeton University,
 Princeton, NJ  08544, USA
\vspace{2mm}
\end{center}

\vspace{3mm}

\begin{center}
{\large Abstract}
\end{center}
The logarithmic running of marginal double-trace operators is
a general feature of 4-d field theories containing scalar
fields in the adjoint or bifundamental representation. 
Such operators provide leading contributions in the large $N$
limit; therefore, the leading terms in their
beta functions must vanish for a theory to
be large $N$ conformal. We calculate the one-loop beta functions in
orbifolds of the ${\cal N}=4$ SYM theory by a discrete subgroup
$\Gamma$ of
the $SU(4)$ R-symmetry, which are dual to string theory
on $AdS_5\times S^5/\Gamma$. We present a general strategy
for determining whether there is a fixed line passing
through the origin of the coupling constant space.
Then we study in detail some classes of non-supersymmetric
orbifold theories,
and emphasize the importance of decoupling
the $U(1)$ factors. Among our examples,
which include orbifolds acting freely on the $S^5$,
we do not find any large $N$ non-supersymmetric
theories with fixed lines passing through the origin.
Connection of these results with closed string tachyon
condensation in $AdS_5\times S^5/\Gamma$ is discussed.

\noindent
\vfil
\begin{flushleft}
May 2005
\end{flushleft}
\vfil
\end{titlepage}

\newpage

\renewcommand{\baselinestretch}{1.1}  

\section{Introduction}

Soon after the AdS$_{d+1}$/CFT$_d$ correspondence was formulated
\cite{jthroat,US,EW} (see \cite{MAGOO,Klebanov:2000me}
for reviews), it was realized that modding out by a discrete
subgroup of the R-symmetry leads to dual pairs with reduced
supersymmetry \cite{KS,LNV}. If we start with the ${\cal N}=4$ SYM
theory in
$d=4$, then a discrete orbifold group $\Gamma \subset SU(2)$ produces
a ${\cal N}=2$ superconformal field theory,
while $\Gamma\subset SU(3)$ produces a
${\cal N}=1$ superconformal gauge theory.
For all other $\Gamma$ the supersymmetry is completely broken,
raising the hope of generating a wide variety of non-supersymmetric
conformal gauge theories. Some support for this was provided
using both string theory \cite{BKV}
and perturbative gauge theory \cite{BJ}
arguments: it was shown that all
correlation functions of single-trace untwisted operators (i.e. the operators
that do not transform under the quantum symmetry $\Gamma$)
coincide in the planar limit with corresponding correlation functions
in the parent ${\cal N}=4$ SYM theory. Therefore, beta functions for
marginal single-trace operators vanish in the large $N$
limit. Concerns were raised, however, about the non-supersymmetric cases
due to the presence of closed string tachyons 
\cite{BKV}.
Nevertheless, the possibility that
non-supersymmetric
orbifold gauge theories are ``large $N$ conformal''
raised interesting prospects of conformal unification without
supersymmetry \cite{FV}.

As briefly mentioned in \cite{BKV, BJ}, double-trace contributions are
not inherited from the parent theory.
An explicit one-loop calculation \cite{TZ}
for the simplest non-supersymmetric
orbifold gauge theory, with $\Gamma=\IZ_2$, revealed the presence
of beta functions for double-trace operators.
The induced double-trace operators were found to be of the form
$O^2$, where $O$ is a twisted ($\IZ_2$ odd) single-trace operator
of bare dimension $2$ (see footnote 11 in \cite{TZ}).
More general concerns about inducing the double-trace
operators were expressed in \cite{Csaki}.
Somewhat later on, the concerns about beta functions for the
double-trace operators were strengthened, since
their presence destroys the scale invariance of the
large $N$ theory \cite{Adams}.\footnote{The role of multi-trace
operators in the AdS/CFT correspondence was examined in a number of
papers, 
starting with \cite{Aharony,Witten, Berkooz}.}
The work of \cite{Adams} draws an important distinction between
the freely acting orbifolds of $AdS_5 \times S^5$ which contain
no tachyons at large radius (strong `t Hooft coupling $\lambda$), and
other orbifolds that do contain tachyons. The case of the
$\IZ_2$ orbifold fits in the context of type 0 string theory
\cite{KT} and therefore contains tachyons. It was speculated
in \cite{Adams} that its Coleman-Weinberg instability
\cite{Coleman}
at weak gauge coupling is related to the tachyonic instability
at strong coupling. One of the results of our paper is
that even freely acting orbifold gauge theories
may be rendered non-conformal
at weak `t Hooft coupling by the flow of certain double-trace couplings.

In recent literature, inspired
by the construction of exactly marginal
deformations in AdS/CFT correspondence \cite{Lunin},
a new proposal has appeared for a non-supersymmetric
gauge theory that is conformal in the large $N$ limit \cite{Frolov}.
This motivates us to revisit the issue of whether there
are non-supersymmetric
orbifolds of the ${\cal N}=4$ theory that are large $N$ conformal
at weak coupling,
which does not seem to be completely settled.
 We study beta functions for double-trace
couplings $f^i$ and show that each such beta function has
3 leading one-loop contributions.
Each one-loop beta function has two zeros, at $f^i = a^i_\pm \lambda$.
If the $a^i_\pm$ are complex then $f^i$ cannot flow to a fixed point,
and the theory is not large $N$ conformal.
But if $a^i_\pm$ are real, then $f^i$ reaches a non-trivial IR stable
fixed point at $f^i = a^i_+ \lambda$.
If all $a^i_+$ are real then we find an interesting weakly coupled
large $N$ CFT, with double-trace operators induced.
But are there such examples?
In this paper we carry out a general one-loop calculation of
induced double-trace operators, and then study in detail
the beta functions for a few classes of examples where
we find that some, but not all, $a^i$ are real.
We do not know a general argument for the non-existence of
perturbatively stable
non-supersymmetric large $N$ CFT's containing scalars in the adjoint
or bifundamental representation;\footnote{
In non-supersymmetric theories containing fields in fundamental
representations there
exist Banks - Zaks fixed points with massless fermions
\cite{BAZA},  and
their recently proposed
generalizations containing also scalar fields \cite{SCHN}. 
These are
isolated fixed points rather than fixed lines. 
} 
a further search for them is certainly warranted.


In the next section we present some considerations
concerning the flow of the double-trace couplings, and in section 3 present
a general formalism for calculating the one-loop beta-functions
in orbifold gauge theories. Then, in section 4 we
consider some simple examples of
non-freely acting orbifolds
whose AdS duals contain tachyons at large radius.
In section 5 we move on to a class of freely acting orbifolds
whose AdS duals do not contain tachyons at large radius.
None of the examples we consider prove to be large $N$ conformal at
weak `t Hooft coupling.
Possible relations between our calculations
and closed string tachyon condensation
are discussed in section 6.

\section{General Considerations}

In the standard convention, the SYM action is
\begin{equation}
S =-\int d^4 x {1\over 2 g_{\scriptscriptstyle YM}^2}
\Tr F_{\mu\nu}^2 + \ldots
\end{equation}
In the `t Hooft large $N$ limit, $g_{\scriptscriptstyle YM}^2 N$ is held fixed;
hence, the coefficient multiplying the single-trace operator
$\Tr F_{\mu\nu}^2$ is of order $N$. In this convention, the $n$-point functions
of single-trace operators are of order $N^{2-n}$.

Now consider gauge theories obtained by orbifolding the parent
$U(N |\Gamma|)$  ${\cal N}=4$ SYM theory by a
discrete symmetry group $\Gamma$.
The single-trace operators come in two types:
the untwisted ones, invariant
under $\Gamma$, and the twisted ones that transform under $\Gamma$.
For example, for $\Gamma= \Z_k$, there are twisted operators $O_l$ that
transform by $e^{2\pi li/k}$ under the generator of
$\IZ_k$. The symmetry prevents such a twisted operator $O$
from being induced in the effective action. However, it does not prevent
the appearance of a double-trace operator $O\bar O$ where $O$ and
$\bar O$ have opposite quantum numbers under $\Gamma$ (e.g., $O_l O_{-l}$ for
$\Gamma= \Z_k$).
Such an operator is of the same order in the large $N$ expansion as
the action $S$, i.e. of order $N^2$. Hence it contributes to observables
in the leading large $N$ limit.

In non-supersymmetric quiver gauge theories,
in general nothing prevents the appearance
of such double-trace operators.
Indeed, one-loop diagrams induce such operators of bare dimension 4
with logarithmically divergent coefficients \cite{TZ,Adams}:
the effective action
at scale $M$ picks up contributions of the form
\footnote{Here and
throughout the paper $\lambda$ denotes the 't~Hooft coupling in the
parent theory: $\lambda=g_{\scriptscriptstyle YM}^2 N|\Gamma|$.}
\begin{equation} \label{induced}
 \int d^4 x \, O \bar O  a_O\lambda^2 \ln (\Lambda/M) \ ,
\end{equation}
where $\Lambda$ is the UV cut-off and $a_O$ is a coefficient determined
through explicit calculations.
Then, perturbative renormalizability
necessitates the addition of trace-squared couplings to the action:
\begin{equation}
\delta S= - \int d^4 x f O\bar O  \ .
\end{equation}
{}From (\ref{induced}) we note that the beta function for $f$
contains a contribution $a_O \lambda^2$.
However, this is not the only contribution to the one-loop beta function.

If the operator $O$ picks up 1-loop anomalous dimension $\gamma_O \lambda$,
then the dimension of $O\bar O$ in the large $N$ limit
is $4 + 2 \gamma_O \lambda $.
This introduces a term $2 \gamma_O f \lambda$ into
the beta function for $f$. Finally, as discussed for example in
\cite{Witten,Berkooz},
there is a positive contribution $v_O f^2$, where
\begin{equation}
\langle O (x) \bar O (0) \rangle = {v_O\over 4\pi^2 |x|^4 }
\ ,
\end{equation}
which comes from fusion of two double-trace operators in the free theory.

Putting the terms together, we find
\begin{equation} \label{genbeta}
M {\partial f\over \partial M} = \beta_f=  v_O f^2 + 2 \gamma_O \lambda f
+ a_O \lambda^2
\ .
\end{equation}
It is crucial that the right hand side  is not suppressed by powers of
$N$; it is a  leading large $N$ effect.
On the other hand, the beta function for $\lambda$ has no such
contribution, due to the theorem of \cite{BKV,BJ}.
Also, counting the powers of $N$ one can show that the double-trace
operators cannot induce any planar beta functions for single-trace couplings.
Therefore,
in the large $N$ limit, $\lambda$ may be dialed as we wish. In particular,
it can be made very small so that the one-loop approximation in
(\ref{genbeta}) is justified.
Then the equation $\beta_f=0$ has two solutions, $f= a_\pm \lambda$,
where
\begin{equation} \label{sol}
a_\pm ={1\over v_O} \left (-\gamma_O \pm \sqrt{ D }\right )\ ,
\qquad D= \gamma_O^2 - a_O v_O
\ .
\end{equation}
If the discriminant $D$ is positive, then
these solutions are real, so that $f$ may flow to the IR stable fixed point
at $f=a_+ \lambda$. This mechanism could make the theory conformal in
an interesting and non-trivial way: in particular the IR theory has
non-vanishing double-trace couplings.\footnote{In actual examples
we will often find that both $a_+$ and $a_-$ are negative, so
that the Hamiltonian
is not obviously bounded from below (to study its positivity
one needs to include both single trace and double-trace
terms quartic in the scalar fields).
However, it is well-known that many large $N$
theories are locally stable for potentials unbounded from below.
Hence, we will not rule out the fixed points with negative
double-trace couplings, although this issue requires further study.}

If $D$ is negative, then (\ref{genbeta}) is positive definite for
real $f$, which signals a violation of conformal invariance.
However, for small $\lambda$, the flow of $f$ is
actually very slow near
the minimum of $\beta_f$ located at $-\gamma_O\lambda/v_O$.
This is evident from the explicit solution of (\ref{genbeta}):
\begin{equation} \label{expsol}
f(M) = - {\gamma_O\lambda\over v_O} +
{b\lambda\over v_O} \tan \left (
{b\lambda\over v_O} \ln (M/\mu) \right )
\ ,
\end{equation}
where we defined $b=\sqrt{-D}$ and chose the boundary condition
$f(\mu)=-d_O\lambda/v_O$. Thus, at weak `t Hooft coupling
$\lambda$, the double-trace parameter $f$ varies very slowly
for a wide range of scales. Still, it blows up
towards positive infinity in the UV at
$M = \mu e^{\pi v_0/(b\lambda)}$ and reaches $-\infty$ in the IR
at
$M = \mu e^{-\pi v_0/(b\lambda)}$. We expect this singular behavior
to be softened by the $1/N$ corrections, which introduce
a positive beta function for $\lambda$ making it approach
zero in the IR.


\section{Double-trace correction for general orbifolds \label{gen_orb}}

In this section we find the beta function of the couplings of the
dimension 4 double-trace twisted operators in general orbifolds of
${\cal N}=4$ SYM theory. There are many gauge fixing choices one can
make. The calculations are substantially simplified if we choose
the dimensional reduction of the ten dimensional background gauge.
We are interested in the 1-loop effective action
for the scalar fields; at this order in perturbation theory the result
can be found by computing the determinant of the kinetic operators.
In Euclidean space and with hermitian generators for the gauge group,
the relevant bosonic terms are
\begin{eqnarray}
S&=&\int d^4x\,{\rm Tr}\!\Big[(\partial_\mu a_\nu)^2 +
(\partial_\mu\varphi^I)^2 \cr
&&~~~~~~~~~
-g^2_{\scriptscriptstyle YM}
[\phi^I,\,a_\mu][\phi^I,\,a_\mu]-
g^2_{\scriptscriptstyle YM}[\phi^I,\,\varphi^J][\phi^I,\,\varphi^J]
-2g^2_{\scriptscriptstyle YM}
[\phi^I,\,\phi^J][\varphi^I,\,\varphi^J]\Big]~~.
\label{gf_action}
\end{eqnarray}
where $\phi^I$ are the background scalar fields and 
we expanded the action to quadratic order in the quantum fields
$a$ and $\varphi$. The fermions couple to the background scalar fields
$\phi$ via Yukawa couplings inherited from minimal couplings in ten
dimensions. 

We will denote by $\Gamma\subset SU(4)$ the orbifold group; if
$\Gamma$ is a subgroup of $SU(2)$ or $SU(3)$ then the
resulting theory preserves ${\cal N}=2$ or ${\cal N}=1$ supersymmetry,
respectively. We will further denote by $g$ the representation of 
the elements of $\Gamma$ in $SU(|\Gamma|N)$,
where it acts by conjugation. The representation of $\Gamma$ in the
spinor and 
vector representation of $SO(4)$ will be denoted by $r_g$ and
$R_{g}$. Presenting the vector representation of $SO(6)$ as the
2-index antisymmetric tensor representation of $SU(4)$, it follows
that $R_g=r_g\otimes r_g$.

We will compute the determinant of the kinetic operator in a general
scalar field background invariant under the orbifold group
\begin{eqnarray}
\phi^I=R_g^{IJ}\,g\,\phi^J\,g^\dagger~~.
\end{eqnarray}
Since we are not considering a nontrivial fermionic background the
contribution of fermionic loops decouple from that of scalar,
vector and ghost loops and can be computed independently. To shorten the
expression of the effective potential let us define:
\begin{eqnarray}
&&A^{IJ|KL}_{g}=
\Tr(\phi^I\phi^Jg^\dagger)
\Tr(\phi^K\phi^Lg)
+
\Tr(\phi^J\phi^Ig)
\Tr(\phi^L\phi^Kg^\dagger)~~.
\end{eqnarray}
We also introduce a notation for the divergent part of a generic
1-loop scalar amplitude:
\begin{eqnarray}
\Div=\int
\frac{d^{4}k}{(2\pi)^4}\,\frac{1}{k^4}=\frac{1}{16\pi^2}\ln
\frac{\Lambda^2}{M^2}~~.
\end{eqnarray}
where $\Lambda$ is the UV cutoff and $M$ is the renormalization
scale. Also, the notation for the contribution to the effective
action will be:
\begin{eqnarray}
\delta S^{\rm nr.~of~loops|nr.~of~traces}_{\rm source ~of
~contribution}
~~.
\end{eqnarray}

\noindent
$\bullet$
Then, the contribution of the fermion loop
to the double-trace part of the effective action is:
\begin{eqnarray}
\label{fermi22}
\delta S^{\rm 1~loop|2~tr}_{\rm
Fermi}=\lambda^2\frac{\Div}{2|\Gamma|} \sum_{g\in \Gamma} \,
\Tr[\gamma^I\gamma^J\gamma^K\gamma^Lr_g]
\left[A^{JI|KL}_{g}+A^{KI|JL}_{g}+ A^{LI|JK}_{g}\right]~~.
\end{eqnarray}
In this form the fermionic contribution to the effective action is
manifestly real. Giving up manifest reality (which of course is
restored in the sum over the orbifold group elements)
it turns out to be possible
to further simplify this expression to:
\begin{eqnarray}
\label{fermi_v1}
&&\delta S^{\rm 1~loop|2~tr}_{\rm Fermi}=\\
&&~~~~~~~~~ =\lambda^2\frac{\Div}{|\Gamma|} \sum_{g\in \Gamma} \,
\Tr[\gamma^I\gamma^J\gamma^K\gamma^Lr_g] \left[
2\Tr(\phi^J\phi^Ig^\dagger)\Tr(\phi^K\phi^Lg) +
\Tr(\phi^K\phi^Ig^\dagger)\Tr(\phi^J\phi^Lg)\right] \nonumber
\end{eqnarray}
In both equations (\ref{fermi22}) and  (\ref{fermi_v1}) $\gamma^I$
denote  the chiral (i.e. $4\times 4$) 6-dimensional Dirac
matrices. In the absence of the orbifold action matrices $r$ the
trace is trivial to compute:
\begin{eqnarray}
\Tr[\gamma^I\gamma^J\gamma^K\gamma^L]=4\left(\delta^{IJ}\delta^{KL}
+\delta^{IL}\delta^{JK}-\delta^{IK}\delta^{JL}\right)~~.
\end{eqnarray}
For general $r_g$ the results for the nonvanishing
components of  $\Tr[\gamma^I\gamma^J\gamma^K\gamma^Lr_g]$ are collected
in the appendix.

\noindent
$\bullet$ The contribution of the vectors, scalars and ghost loops
to the  double-trace part of the effective action has the following
expression:
\begin{eqnarray}
\label{vs22}
\delta S^{\rm 1~loop|2~tr}_{\rm Bose,\,ghost}
\!\!&=&\!\!
-\lambda^2\frac{\Div}{2|\Gamma|}
\sum_{g\in \Gamma}
\Big\{
~~
\left(\Tr[R_g]+2\right)\left(\Tr(\phi^2 g^\dagger)\Tr(\phi^2 g)
+2\Tr(\phi^I\phi^J g^\dagger)\Tr(\phi^J\phi^I g) \right) \cr
&&+\vphantom{\Big|}
2(R_g^{KQ}+(R_g^{-1}){}^{KQ} )\Tr\left([\phi^I,\,\phi^Q]g^\dagger\right)
\Tr\left([\phi^I,\,\phi^K]\gamma\right)\\
&&-\vphantom{\Big|}
2\left(\delta^{KI}(R_g^{-1}){}^{PQ}+\delta^{PQ}R_g^{KI}
-2\delta^{PQ}\delta^{KI}
\right)\Tr\left(\phi^P\phi^Qg^\dagger\right)
\Tr\left(\phi^K\phi^Ig\right)~~\Big\}
\nonumber
\end{eqnarray}
One may derive this directly in terms of Feynman diagrams or by
extracting the double-trace part of the determinant of the kinetic
operator for the quantum fields in (\ref{gf_action}).  It is trivial
to check that in the absence of the orbifold projection
\begin{eqnarray}
\delta S^{\rm 1~loop|2~tr}_{\rm Bose,\,ghost}
+\delta S^{\rm 1~loop}_{\rm Fermi}=0
\end{eqnarray}
in agreement with the theorem of \cite{BKV,BJ}.

We see that double-trace operators
made out of twisted single-trace operators are
generated at 1-loop. Therefore they must be added to the tree level
action. The precise form of the deformation depends on the specific
orbifold. Also, whenever possible, it is useful to reorganize the
operators being generated in terms of operators with definite scaling
dimension. For the purpose of illustration let us consider the
deformation
\begin{eqnarray}
\delta_{\rm 2~trace}S={1\over
2}\sum_{g\in\Gamma}\,f_g\,O_g^{IJ}O_{g^\dagger}^{JI} ~~~~{\rm
with}~~~~O_g^{IJ}=\Tr(g\phi^I\phi^J)~~.
\label{def}
\end{eqnarray}
This modifies the kinetic operator by adding
\begin{eqnarray}
\sum_{g\in\Gamma}
\left(L(g)O_{g^\dagger}^{JI}+R(g)O_{g^\dagger}^{IJ}\right)
\end{eqnarray}
where $L(\cdot)$ and $R(\cdot)$ are the left- and right-multiplication
operators, respectively, and brings the following
additional contributions to the effective
action:
\begin{eqnarray}
\label{2trcontrib}
&&\delta S^{\rm 1~loop|2~tr}_{\rm 2~trace}
=-\frac{\Div}{|\Gamma|}\Big\{
\sum_g f_g\left(\frac{1}{|\Gamma|}\sum_{\tilde g}
f_{{\tilde g} g^\dagger {\tilde g}^\dagger}\right)O_g^{IJ}O_{g^\dagger}^{JI}\\
\!\!&+&\!\!\lambda
\sum_{g\in\Gamma} f_g O^{JI}_{g^\dagger}
\left[
4\delta^{I{\hat I}}\delta^{J{\hat
J}}+\left(\delta^{IJ}+R_g^{JI}\right)\delta^{{\hat I}{\hat
J}}-2\left(\left(R_g^{-1}\right){}^{I{\hat I}}\delta^{J{\hat J}} +
\delta^{I{\hat I}} \left(R_g^{-1}\right){}^{J{\hat J}}\right)
\right]O_g^{{\hat I}{\hat J}}\Big\}\nonumber
\end{eqnarray}
One may recognize the bracket on the second line as the 1-loop
dilatation operator acting on twisted operators.


\section{Examples of non-freely acting orbifolds}

In this section we review and extend earlier analysis of orbifold
field theories in which the action of the orbifold group on the
$R$-symmetry representation possesses fixed points. Quite generally,
such an orbifold action yields in the daughter theory fields in the
adjoint representation of all the gauge group factors.

On the string theory side, this translates into the existence of fixed
points of the action of the orbifold group on the five-sphere. For
non-supersymmetric actions (such as those we are interested in) it
follows that some of the string theory excitations are tachyonic.
We will eventually show that such tachyons manifest themselves
in the weakly coupled gauge theory.

\subsection{A Non-Supersymmetric $\IZ_2$ Example \label{Z2example}}

In this subsection
we discuss the $\IZ_2$ orbifold theory which arises on the stack of
$N$ electric and $N$ magnetic D3-branes of type 0B theory \cite{KT}.
This is the $SU(N) \times SU(N)$ gauge theory coupled to six adjoint
scalars
$\Phi^I$ of the first gauge group, six adjoint scalars $\tilde \Phi^I$
of the second gauge group, 4 fermions in
$({\bf N}, {\bf \bar N})$
and 4 fermions in
$({\bf \bar N}, {\bf N})$. This theory has global $SO(6)$ symmetry.

The one-loop calculation of \cite{TZ} reveals the following double-trace
terms induced in the effective Lagrangian:
\begin{equation}
\delta{\cal L}^{\it eff}=
{\lambda^2\over\pi^2}\ln\left(\frac{\Lambda}{M}\right)
\left(O^{\langle IJ\rangle} O^{\langle IJ\rangle} +{2\over 3} O^2 \right) \ ,
\end{equation}
where
\begin{equation}
O^{\langle IJ\rangle}= \Tr(\Phi^I\Phi^J -{1\over 6}\delta^{IJ}\Phi^K\Phi^K)
-\Tr (\tilde\Phi^I\tilde\Phi^J -{1\over 6}\delta^{IJ} \tilde\Phi^K
\tilde\Phi^K)\equiv \Tr( g\phi^I\phi^J) -\frac{1}{6}\delta^{IJ}
\Tr( g\phi^K\phi^K)
\end{equation}
transform in the ${\bf 20}$ of $SO(6)$, while
\begin{equation}
O =
\Tr\Phi^I\Phi^I
-\Tr \tilde\Phi^I\tilde\Phi^I\equiv \Tr( g\phi^I\phi^I)
\end{equation}
is an $SO(6)$ singlet. Here the matrix $g$ represents the $\IZ_2$
orbifold group on the gauge degrees of freedom: $g={\rm diag}(\id_N,-\id_N)$.
We are thus forced to introduce coupling constants $f_{\bf 20}$ and
$f_{\bf 1}$,
through
\begin{equation}
\delta {\cal L}^{\it tree} = - f_{\bf 20} O^{\langle IJ\rangle} O^{\langle IJ\rangle} -
f_{\bf 1} O^2 \ .
\end{equation}

With our conventions (the normalization of the scalar kinetic term
is twice the usual), the free scalar Euclidean two-point function is
\begin{equation}
\langle \Phi^I (x) \Phi^J (0) \rangle =\delta^{IJ} {1\over 8\pi^2 |x|^2 }\ .
\end{equation}
Then we find
\begin{equation}
\langle O (x) O (0) \rangle = {v_{\bf 1}\over 4\pi^2 |x|^4 }\ ,
\qquad v_{\bf 1}= {3\over 4\pi^2}
\ ,
\end{equation}
\begin{equation}
\langle O^{\langle IJ\rangle} (x) O^{\langle KL\rangle} (0) \rangle = (\delta^{IK}
\delta^{JL} + \delta^{IL} \delta^{JK}- {1\over
3}\delta^{IJ}\delta^{KL}) {v_{\bf 20}\over 4\pi^2 |x|^4 } \ ,
\qquad v_{\bf 20}= {1\over 8\pi^2}\ .
\end{equation}
The one-loop anomalous dimension coefficients are
\begin{eqnarray}
\gamma_{\bf 1}= {3\lambda\over 8\pi^2}\ ,\qquad
 \gamma_{\bf 20}= 0~~.
\end{eqnarray}
They can be obtained from the corresponding quantities in
the ${\cal N}=4$ SYM theory (see, for instance, \cite{Bianchi})
by interpreting $\lambda$ as the 't~Hooft coupling in
the parent theory and introducing an additional factor
of $1/|\Gamma|$ or by diagonalizing the dilatation operator
written out explicitly in the appendix. Hence, we find
\begin{equation}
\beta_{\bf 20} = v_{\bf 20} f_{\bf 20}^2 + {\lambda^2\over \pi^2}
\ ,
\qquad
\beta_{\bf 1} =  v_{\bf 1} f_{\bf 1}^2 +
{3\over 4 \pi^2} \lambda f_{\bf 1} + {2\lambda^2\over 3\pi^2}
= {3\over 2 \pi^2}\left ( \frac{1}{4}f_{\bf 1}^2 +
\frac{1}{2} \lambda f_{\bf 1} +
{4\over 9} \lambda^2\right ) \ .
\end{equation}
Neither $\beta_{\bf 20}$ nor $\beta_{\bf 1}$ have real zero's:
they are positive definite for real couplings.
Hence, the double-trace couplings $f_{\bf 20}$ and $f_{\bf 1}$
flow from large positive values in the UV to large negative
in the IR. Thus, the $\IZ_2$
orbifold theory is not large $N$ conformal: there are
$\IZ_2$ odd single-trace operators
and $\IZ_2$ even double-trace operators whose
correlators do not respect conformal
invariance.

\subsection{A Non-supersymmetric $\IZ_3$ orbifold}

As usual, we start with a ${\cal N}=4$ supersymmetric
$SU(3N)$ gauge theory and apply a projection.
We take the generator of the $\IZ_3$ group to act on the fundamental
representation of $SU(4)$ as
\begin{eqnarray}
r={\rm diag} (e^{i\alpha_3}, e^{i\alpha_3},
e^{-i\alpha_3}, e^{-i\alpha_3} )\ ,~~~~\alpha_3=\frac{2\pi}{3}~~.
\end{eqnarray}
The action on the fundamental representation of $SO(6)$ is
\begin{eqnarray}
R={\rm diag} (1,1,e^{2i\alpha_3}, 1,1, e^{-2i\alpha_3})
\ .\end{eqnarray}
Thus, the $\IZ_3$
orbifold acts on only one of the three complex coordinates.
Closed string tachyon condensation in the $\IC/\IZ_3$ case,
and in the generalization to $\IC/\IZ_k$ discussed in
the Appendix \ref{SO4}, was studied in many papers starting with
\cite{ADPS} (for reviews see \cite{Martinec,HEMT}).
Our strategy will be to place a stack of $N k$ D3-branes at the tip of
the cone, and to study RG flows in the resulting
$U(N)^k$ gauge theory, and we will suggest their connection with
tachyon condensation.

As usual in orbifold field theories,
we keep only fields invariant under this operation, together with
$A\to g A g^\dagger$, where
$$ g= {\rm diag} (\id_N, e^{i\alpha_3} \id_N, e^{-i\alpha_3}
\id_N)~~.
$$
where $\id_N$ denotes the $N\times N$ identity matrix.

We end up with a $U(N)^3/U(1)$ gauge theory
(the untwisted $U(1)$ decouples) described by a quiver
diagram with 3 vertices. This theory has no supersymmetry but possesses
$SO(4)\sim SU(2)\times SU(2)$ global symmetry.
At each vertex of the quiver, there are 4 adjoint scalar fields,
transforming as a vector of $SO(4)$, $\Phi_i^\mu$. Here
$\mu=1,2,3,4$ is the $SO(4)$ index, and $i=1,2,3$ labels the
vertex of the quiver. There are also $SO(4)$ singlet bifundamental
scalars $\Phi_{ij}$ with $i\neq j$.
In the fermionic sector, we find 3 doublets of the first $SU(2)$,
$\psi_{12}^a$, $\psi_{23}^b$, $\psi_{31}^c$, with $a,b,c=1,2$;
and 3 doublets of the second $SU(2)$,
$\psi_{21}^{\dot a}$,
$\psi_{32}^{\dot b}$, $\psi_{13}^{\dot c}$.
The Yukawa couplings include terms of the type
$$\Phi_1^\mu \sigma^\mu_{a\dot b} \psi_{12}^a \psi_{21}^{\dot b}$$
and also terms of the type
$$ \epsilon_{ab} \psi_{12}^a \psi_{23}^b \Phi_{31}\ .
$$


First, let us classify the scalar operators that may appear in
the induced marginal double-trace operators.
The operators built of the adjoint scalars can be combined into
traceless symmetric
tensors in the ${\bf 9}$ of $SO(4)$, and also into the singlet of
$SO(4)$. The former have the form
\begin{equation}
O_\pm^{\langle \mu\nu \rangle }= O_1^{\langle\mu\nu\rangle}+
\exp (\pm i \alpha_3) O_2^{\langle \mu\nu\rangle}+
\exp (\mp i \alpha_3) O_3^{\langle \mu\nu\rangle}
\ ,
\end{equation}
where
\begin{equation}
 O_i^{\langle \mu\nu\rangle} = \Tr (\Phi_i^\mu \Phi_i^\nu - {1\over 4}
\delta^{\mu\nu} \Phi_i^\kappa \Phi_i^\kappa)
~~~~~~~~
O_\pm^{\langle \mu\nu \rangle }=\Tr (g^{\pm 1}\phi^{\mu}\phi^\nu)
-\frac{1}{4}\delta^{\mu\nu}\,\Tr(g^{\pm 1}\phi^{\kappa}\phi^\kappa)
\ ,
\end{equation}
while the latter are
\begin{equation}
O_\pm =\sum_{\kappa=1}^4\Tr (g^{\pm 1}\phi^\kappa \phi^\kappa)
= \Tr \Phi_1^2 + \exp (\pm i \alpha_3)  \Tr \Phi_2^2 + \exp
(\mp i \alpha_3) \Tr \Phi_3^2 \ .
\end{equation}
Additionally, there are $SO(4)$ singlet operators made of
the bifundamental scalars,
\begin{eqnarray}
A_{\pm}=e^{\pm i\alpha_3}
\Tr(g^{\pm1}\phi^3\phi^{\bar 3})=\sum_{k=1}^3 \Phi_{k,k+1}\Phi_{k+1,k}
e^{\pm i \alpha_3 k}
~~~~{\rm where}~~~~
\Phi_{kl}^\dagger =\Phi_{lk} \ ,
\end{eqnarray}
and $k+3$ is identified with $k$.

The operators $O_\pm$ and $A_\pm$ mix under RG flow; their anomalous
dimension matrix is:
\begin{eqnarray}
\Delta
\begin{pmatrix}
O_\pm\cr A_\pm
\end{pmatrix}
=\frac{1}{48\pi^2}\delta
\begin{pmatrix}
O_\pm\cr A_\pm
\end{pmatrix}
\equiv
\frac{1}{48\pi^2}
\begin{pmatrix}
\delta^{OO}&\delta^{OA}\cr
\delta^{AO}& \delta^{AA}
\end{pmatrix}
\begin{pmatrix}
O_\pm\cr A_\pm
\end{pmatrix}
=
\frac{\lambda}{48\pi^2}
\begin{pmatrix}
\hphantom{-}8&-8\cr
-1& \hphantom{-}7
\end{pmatrix}
\begin{pmatrix}
O_\pm\cr A_\pm
\end{pmatrix}
\label{anomdimZ3}
\end{eqnarray}

The $\IZ_3$ permutation symmetry, and the $SO(4)$ symmetry
imply that the double-trace operators must involve combinations
like
\begin{equation}
O_+^{\mu\nu} O_-^{\mu\nu}=
\sum_{i=1}^3 O_i^{\mu\nu} O_i^{\mu\nu}
-O_1^{\mu\nu} O_2^{\mu\nu}
-O_2^{\mu\nu} O_3^{\mu\nu}
-O_1^{\mu\nu} O_3^{\mu\nu}
\ .
\end{equation}
This is a good check on our calculations since such combinations
emerge only after we sum over the gauge field, scalar and
fermion loops.

Explicit calculation shows that there are several combinations which
are being generated at one-loop level and must therefore be added as
tree-level deformations of the original action. They are:
\begin{eqnarray}
\delta_{\rm 2~trace}{\cal L}^{\rm tree}=
 f_{{\bf 9},1}O^{\langle \mu\nu \rangle}_+O^{\langle \mu\nu \rangle}_-
+f_{{\bf 1},1}O_+O_-
+f_{(3),1}A_+A_-
+f \left(A_+O_-+A_-O_+\right)
\end{eqnarray}
In the following we will keep the anomalous dimension matrix
(\ref{anomdimZ3}) nondiagonal. This leads to non-diagonal beta
functions, but avoids explicitly using the matrix diagonalizing
the anomalous dimension matrix.

Specifying the results from the Appendix \ref{SO4} 
to $\IZ_3$ we find that, in
the presence of the deformation, the
double-trace part of the one-loop effective potential is:
%
%
%
%
\begin{eqnarray}
\label{effso4}
\delta
S^{\rm 1\,{\scriptscriptstyle loop}|2\,tr}_{\rm \scriptscriptstyle{b,gh,f}}
=-\frac{9\lambda^2\ln(\Lambda^2/\mu^2)}{32\pi^2|\Gamma|}
\Bigg[
\,4\,O_+^{\langle \mu\nu\rangle}O_{-}^{\langle \mu\nu\rangle}
+3\,O_+O_{-}
+18A_+A_{-}
-6\left(O_+A_{-}
+O_{-}A_{+}\right)
\Bigg]
\end{eqnarray}
%
%
%
%
\begin{eqnarray}
\delta S^{\rm 1\,{\scriptscriptstyle loop}|2\,tr}_{\rm 2\,trace}
\!\!\!\!&=&\!\!\!\!
-\lambda^2\frac{\ln(\Lambda^2/\mu^2)}{32\pi^2|\Gamma|}\Bigg[
{f^2_{{\bf 9},1}} O_+^{\langle ab \rangle} O_{-}^{\langle ab
\rangle} + \left[ 4{f^2_{{\bf 1},1}}
+{\textstyle{\frac{1}{2}}}f^2 +
2f_{{\bf 1},1} \delta^{OO}
+2f\delta^{AO}\right]O_+ O_{-} \cr
&&~~~~~~~~~~\vphantom{\Big|}
+
\left[{\textstyle{\frac{1}{2}}}{f^2_{(3),1}}
+
4f^2
+
2f_{(3),1}\delta^{AA}
+2f\delta^{OA}\right] A_+ A_{-}\\
&&~~~~~~~~~~\vphantom{\Big|}
+
\left[f\left(
4f_{{\bf 1},1}+
{\textstyle{\frac{1}{
2}}}f_{(3),1}+\delta^{OO}+\delta^{AA}\right)
+
f_{{\bf 1},1}\delta^{OA}+f_{(3),1}\delta^{AO}\right]O_+ A_{-}\cr
&&~~~~~~~~~~\vphantom{\Big|}
+
\left[f\left(
4f_{{\bf 1},1}+
{\textstyle{\frac{1}{
2}}}f_{(3),1}+\delta^{OO}+\delta^{AA}\right)
+
f_{{\bf 1},1}\delta^{OA}+f_{(3),1}\delta^{AO}\right]O_- A_{+}
~~~~\Bigg] \nonumber
\end{eqnarray}

The five beta functions are therefore:
\begin{eqnarray}
\beta_{{\bf 9},1}&=&
\frac{1}{48\pi^2}\left[36\lambda^2+f^2_{{\bf 9},1} \right]
\cr
\beta_{{\bf 1},1}&=&
\frac{1}{48\pi^2}\left[27\lambda^2+
4{f^2_{{\bf 1},1}}
+{\textstyle{\frac{1}{2}}}f^2
+
16\lambda f_{{\bf 1},1}
-2\lambda\,f
\right]
\cr
\beta_{(3),1}&=&
\frac{1}{48\pi^2}\left[162\lambda^2+
{\textstyle{\frac{1}{2}}}{f^2_{(3),1}}+
4f^2+
14\lambda f_{(3),1}
-16\lambda\,f
\right]
\cr
\beta_{f}&=&
\frac{1}{48\pi^2}\left[-54\lambda^2+f\left(
4f_{{\bf 1},1}+
{\textstyle{\frac{1}{
2}}}f_{(3),1}+15\lambda
\right)
-8\lambda f_{{\bf 1}, 1}-\lambda f_{(3),1}
\right]
\end{eqnarray}

These expressions may seem quite opaque; it is however relatively easy
to analyze them and find that no real values for the couplings $f$
lead to vanishing of all four beta functions. This is quite obvious
for $\beta_{{\bf 9},1}$, which corresponds to operators with vanishing
one loop anomalous dimension.
In the next section we show how this generalizes to any
non-freely acting $\IZ_k$ orbifold.

\subsection{General non-freely acting $\IZ_k$ orbifolds}

In both examples discussed above we found that there is no weakly
coupled fixed point of the RG flow. We will now show that this is in
fact a general property of  $\IZ_k$ orbifolds with fixed points by
identifying operators whose beta function does not vanish for any value
of the coupling constants.

First of all, let us classify all possible representations of
$\IZ_k$ embedded in $SU(4)\simeq SO(6)$. Through a unitary
transformation, its only nontrivial generator
can be brought to a diagonal form
\begin{eqnarray}
g=
\left(%
\begin{array}{cccc}
  e^{in_1\alpha_k} & 0 & 0 \\
  0& e^{in_2\alpha_k} & 0 & 0 \\
  0 & 0 & e^{in_3\alpha_k} &0 \\
0 & 0 & 0 & e^{in_4\alpha_k}
\end{array}%
\right),~~~~~~~~\alpha_k={2\pi\over k}
\end{eqnarray}
with a constraint $n_1+n_2+n_3+n_4=0$. So the
representation is specified by three integers $n_i\in \IZ~
mod~k$. Since the fundamental representation of $SO(6)$ is
isomorphic to the two-index antisymmetric representation of $SU(4)$,
it follows that the action of $\IZ_k$ on the fundamental
representation of $SO(6)$ is also specified by three integers and
their negatives $\{m\}=\{n_i+n_j|i\neq j,~ i,j=1,2,3, 4\}$. These are
the weights of the complex scalar fields and their conjugates under
this action of $\IZ_k$.

The integers $m$ vanish in pairs and the $\IZ_k$-invariant
subspace of $\IR^6$ is always even-dimensional. We will focus in
this section on representations with at least one vanishing $m$,
that is we will choose
\bea
n_1=-n_2=n'
~~~~~~~{\rm and}~~~~~~~~~
n_3=-n_4=n''~~.
\eea
Let us denote by $2l$ the number of vanishing weights.
Then, $SO(2l)\in SO(6)$ is the remaining unbroken global symmetry
of the theory. We will denote by $\mu,\nu,..$ the indices along
$\IZ_k$-invariant directions and by $i,j,..$ all the others directions.

Our logic will be the same as in the examples discussed before: we
will focus on the symmetric $SO(2l)$ traceless operator
\be
O_q^{\langle \mu\nu\rangle}=\Tr(g^q\phi^\mu\phi^\nu)-{\delta^{\mu\nu}\over
2l}\Tr(g^q\phi^\kappa\phi^\kappa)
\label{def_nf}
\ee
and the beta-function for the corresponding coupling
constants\footnote{The reason for this particular form of the
tree-level deformation
is that it yields a uniform expression for all beta functions,
including the operators with charge $q=k/2=[k/2]$.}
\begin{eqnarray}
\delta{\cal L}^{\it tree}= \frac{1}{2}\sum_{q=1}^{k-1}\,f_q
O_q^{\langle \mu\nu\rangle}O_{-q}^{\langle \mu\nu\rangle} ~~~{\rm with}~~~ f_q=f_{k-q}
\end{eqnarray}
When computing the beta function we have to remember to take into
account this overcounting of operators.

The only other twisted operators containing fields from the invariant
subspace are similar to the Konishi operator in the parent theory;
explicitly, they are
\be
O_q=\sum_\kappa \Tr(g^q\phi^\kappa\phi^\kappa)
\ee
where the sum runs only over the invariant subspace. The only other
potential candidate
\be
\Tr(g^q\phi^\mu\phi^i)
\ee
vanishes identically. This can be proven quite easily by moving one
factor of $g$ past both $\phi^\mu$ and $\phi^i$ and then using the
cyclicity of the trace. This operation yields a nontrivial phase
proportional to the charge of $\phi^i$. The other twisted operators
are
\begin{eqnarray}
\Tr(g^q\phi^i\phi^{\jbar})~~.
\end{eqnarray}

This spectrum clearly implies that the deformation (\ref{def_nf}) is
closed in the sense that the beta functions for the couplings $f_q$
does not receive contributions linear in other couplings. Indeed, all
correlation functions
\be
\langle O_q^{\langle \mu\nu\rangle}O_{-q}\rangle=0
\ee
since there is no traceless symmetric $SO(2l)$ invariant tensor.

{}From the general expressions listed in Appendix A, it is easy to see
that $O^{\mu\nu}_q$ has vanishing one-loop anomalous dimension, so
there is no contribution of the type $\lambda f_q$ to the
corresponding beta function. It follows therefore that
there are only two relevant contributions: from
$\langle O_q^{\mu_1\nu_1}O_{-q}^{\mu_2\nu_2}\rangle$
and from the one-loop renormalization of the bare action.
The former is always positive  \cite{Witten}. It is in fact easy
to calculate this coefficient at one loop level. Its corresponding
contribution to the beta function is
\begin{eqnarray}
 { f_q^2\over 16\pi^2\,k}>0~~.
\end{eqnarray}
The later contribution, from the one loop renormalization of the
bare action
is
\begin{equation}
\delta
S=-\frac{\lambda^2}{\pi^2\,k}\ln\left(\frac{\Lambda}{M}\right) ~
\sum_{q=1}^{k-1} \sin^2({n'\alpha q\over 2})\sin^2({n''\alpha
q\over 2})~ O_q^{\mu\nu}O_{-q}^{\mu\nu}
\end{equation}
and is also positive.Thus, the beta function for the operators
(\ref{def_nf}) is
\be
\beta_{q}=
{2\lambda^2\over \pi^2\,k} \sin^2({n'\alpha_k\over 2})
\sin^2({n''\alpha_k\over 2}) + { f_q^2\over 16\pi^2 k}
\ee
and is always non-vanishing.


\section{A class of orbifolds with $SU(3)$ symmetry}

\bigskip

In the previous section we saw that, quite generally,
non-supersymmetric orbifolds that are not freely acting do not correspond
to weakly coupled large $N$ CFT's.
What about freely-acting orbifolds?
A motivation for studying them \cite{Adams} is that,
since they have no fixed points,
the twisted sector strings are stretched to length
of order $R\sim \lambda^{1/4} \sqrt{\alpha'}$ and therefore
are not tachyonic at large `t Hooft coupling $\lambda$.
The corresponding fields in $AdS_5$ have $m^2\sim \sqrt{\lambda/\alpha'}$.
Such fields are dual to the twisted single-trace
operators in the orbifold gauge theory,
which are charged under the quantum symmetry $\Gamma$ \cite{Gukov}.
By the AdS/CFT correspondence, at large $\lambda$ such operators have dimensions
of order $\sqrt \lambda$ and are highly irrelevant. Hence, the AdS/CFT
correspondence suggests that there is a fixed line at large $\lambda$, but
that instabilities may set in for small $\lambda$. Motivated by this,
we carry out the small $\lambda$ (one-loop) analysis
for a class of freely acting $\IZ_k$ orbifold
gauge theories which possess a global $SU(3)$ symmetry (further details
may be found in the Appendix B).


As before, the starting point is ${\cal N}=4$ SYM theory with gauge group
$U(kN)$;  let us parametrize $\IZ_k=\{g^n|n=0,\dots, k-1,g={\rm
diag}(1,\omega_k,\dots \omega_k^{k-1})\}$ where $\omega_k$ is the first
$k$-th root of unity
\begin{eqnarray}
\omega_k=e^{i\alpha_k}\ ,~~~~~~~\alpha_k=\frac{2\pi}{k}~~.
\end{eqnarray}
To preserve $SU(3)$, we choose the following
action of $\IZ_k$ in the fundamental representation of $SU(4)$:
\begin{eqnarray}
r(g^n)={\rm diag}(\omega_k^n,\omega_k^n,\omega_k^n,\omega_k^{-3n})\ .
\label{spinSU3}
\end{eqnarray}
which yields the action in the vector representation of $SO(6)$
\begin{eqnarray}
R(g^n) ={\rm diag}(\omega_k^{2n},\omega_k^{2n},\omega_k^{2n},
\omega_k^{-2n},\omega_k^{-2n},\omega_k^{-2n})
\label{vecSU3}
\end{eqnarray}

We end up with a $U(N)^k/U(1)$ gauge theory
described by a quiver
diagram with $k$ vertices. This theory has no supersymmetry
but possesses $SU(3)$ global symmetry. On the edges around the boundary
of the quiver there  are $k$ $SU(3)$ triplets of chiral fermions,
$\psi^i_{m,m+1}$, where $m$ is identified with $m+k$.
There are also $k$ $SU(3)$ singlet chiral fermions
$\chi_{m, m-3}$.
In the scalar sector, we find
$k$ complex $SU(3)$ triplets, $\Phi^i_{m,m+2}$.
Since there are no adjoint scalars, the simplest Coleman-Weinberg
potential of the type considered in \cite{TZ,Adams} cannot be
generated. However, these models, like all other orbifolds, contain
single-trace operators quadratic in the scalar fields.
Therefore, there is a possibility of inducing beta-functions
for double-trace operators made out of twisted single-trace operators.

The spectrum of invariant fields under this combined action allows the
construction of the following independent twisted operators:
\begin{eqnarray}
O^{i\jbar}_n=\Tr(g^n\phi^i\phi^\jbar)
~~~~~~
O_n=\sum_{i=1}^3\Tr(g^n\phi^i\phi^\ibar)
~~~n=1,\dots,\left[\frac{k}{2}\right]
\end{eqnarray}
These operators mix under 1-loop scale transformations. Using the
dilatation operator spelled out in the appendix it is easy to identify
the operators with definite scaling dimension:
\begin{eqnarray}
\begin{array}{c|c}
{\rm Operator} & {\rm anomalous ~dimension}\cr\hline\cr
O^{\langle i\jbar
\rangle}_n=\Tr(g^n\phi^i\phi^\jbar)-\frac{1}{3}\eta^{i\jbar}\,O_n
&  \frac{\lambda}{16\pi^2 k}\left[8\sin^2 (n\alpha_k)\right]
 \cr  & \cr\hline\cr
O_n=\sum_{i=1}^3\Tr(g^n\phi^i\phi^\ibar) &
\frac{\lambda}{16\pi^2 k}\left[2(5+\cos (2n\alpha_k) )\right]
\end{array}
\label{anomdimSU3}
\end{eqnarray}
These operators transform in the octet and singlet representation of
the $SU(3)$ global symmetry group.


It is important to make a distinction between $\IZ_{2m+1}$
and $\IZ_{2m}$. Only the former is freely acting.
The latter has a $\IZ_2$ subgroup
$\{1,g^m\}$ which does not act freely since
$r(g^m)= {\rm diag} (-1,-1,-1,-1)$.
This suggests that the
corresponding orbifold theory is {\it similar to} a $\IZ_m$
orbifold of the $\IZ_2$ theory discussed in
section \ref{Z2example}.\footnote{A detailed
discussion of this effect for the $\IZ_4$ orbifold, corresponding to
$m=2$, may be found in \cite{KNS}.}
The operators carrying $m$ units of charge, while in the twisted
sector of the $\IZ_{2m}$
orbifold, are in fact inherited from the initial $\IZ_2$ orbifold.
Thus, through the inheritance principle, the ${\cal O}(\lambda^2)$
contribution to their beta functions should
be related by a rescaling of the coupling to the
beta functions we found in \S \ref{Z2example}.\footnote{We will also
find additional numerical factors related to the different detailed
structure of the operators generated at one-loop level. However, the
physics following from this beta function is universal.}.
This identification suggests that on the string theory side the
$\IZ_{2m}$ orbifolds with the action (\ref{spinSU3})-(\ref{vecSU3})
contain
tachyons of the type 0B theory.


\subsection{A non-supersymmetric $\IZ_5$ orbifold}

\begin{figure}[ht]
\begin{center}
\epsfig{file=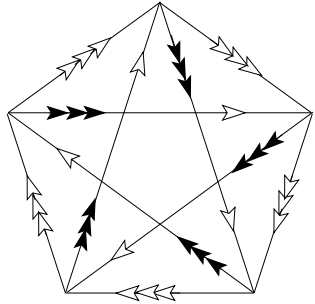,height=4cm}
\caption{ The quiver diagram for the non-supersymmetric freely acting
$\IZ_5$ orbifold. \label{Z5quiver}}
\end{center}
\end{figure}

This is the smallest non-supersymmetric freely acting orbifold.
In this case, the quiver summarizing the field content described above
is shown in figure \ref{Z5quiver}. The white arrows denote fermions
and the black arrows denote scalars.
There are five $SU(3)$ triplets of chiral fermions,
$\psi^i_{k,k+1}$, on the boundary of the quiver as well as
five $SU(3)$ singlet chiral fermions $\chi_{k, k+2}$ and
five complex $SU(3)$ triplets, $\Phi^i_{k,k+2}$
corresponding to the edges of the star.

The explicit form of the $SU(3)$ octets (\ref{anomdimSU3}) is
\begin{equation}
O_n^{\langle i\jbar \rangle}\equiv \Tr(g^n \phi^i\phi^\jbar) -
\frac{1}{3}\eta^{i\jbar} \Tr(g^n \phi^k\phi^{\bar k})=\sum_{k=1}^5
\left( \Phi^i_{k,k+2}\Phi^{\jbar }_{k+2,k} -{1\over 3}\eta^{i\jbar }
\Phi^l_{k,k+2}\Phi^{\bar l}_{k+2,k} \right)
e^{ i n\alpha (k-1)}~~,
\end{equation}
where $n$ assumes values $-2,-1,0,1,2$. The deformation of the tree
level Lagrangian is:
\begin{eqnarray}
\delta_{\rm 2~trace}{\cal L}=
 f_{{\bf 8},1}O_1^{\langle i{\jbar }\rangle}
    O_{-1}^{\langle j\ibar\rangle}
+f_{{\bf 8},2} O_2^{\langle i \jbar\rangle}
    O_{-2}^{\langle j{\ibar }\rangle}
+f_{{\bf 1},1}O_1
    O_{-1}
+f_{{\bf 1},2} O_2
    O_{-2}
\end{eqnarray}
Putting this together with the general expression in \S \ref{gen_orb}
yields the beta functions for the four twisted couplings $f$:
\begin{eqnarray}
\beta_{{\bf 8},1}&=&
\frac{1}{160\pi^2 }
\left[
\left(f_{{\bf 8},1} + 2\left(5+\sqrt{5}\right) \right)^2
-40(\sqrt{5}+1) {\lambda^2} \right]\cr
\beta_{{\bf 8},2}&=&
\frac{1}{160\pi^2 }
\left[
\left(f_{{\bf 8},2} + 2 (5-\sqrt{5})\right)^2
+40(\sqrt{5}-1) {\lambda^2} \right]\cr
\beta_{{\bf 1},1}&=&
\frac{3}{160\pi^2 }
\left[
\left(f_{{\bf 1}, 1} + \frac{1}{3}\left(19-\sqrt{5}\right)\right)^2
-\frac{16}{9}(7\sqrt{5}+1) {\lambda^2} \right]\cr
\beta_{{\bf 1},2}&=&
\frac{3}{160\pi^2 }
\left[
\left(f_{{\bf 1}, 2} + \frac{1}{3}\left(19+\sqrt{5}\right)\right)^2
+\frac{16}{9}(7\sqrt{5}-1)  {\lambda^2} \right]
\end{eqnarray}

We observe that the beta functions for the singly-charged
octet and singlet operators have nontrivial zeroes at real values of the
coupling, while the coupling constants of the doubly-charged operators
do not.

\subsection{The ${\cal N}=1$ supersymmetric $\IZ_3$ orbifold}

The smallest quiver gauge theory in the class (\ref{spinSU3}) is
the $\IZ_3$, which has $\omega_3= e^{2\pi i/3}$.
In this case, one of the eigenvalues of $r(g)$ equals $1$, hence the
orbifold preserves ${\cal N}=1$ supersymmetry \cite{KS,LNV}.
One finds $U(N)^3/U(1)$ supersymmetric gauge theory coupled to three
bifundamental chiral superfields on each edge of the triangular quiver
diagram. While outside the main goal of this section, the analysis of this
theory exposes important subtleties.

The single-trace operators quadratic in the scalar fields
again break up into octets
\begin{equation}
O_n^{\langle i\jbar \rangle} =\sum_{k=1}^3
\left( \Phi^i_{k,k-1}\Phi^{\jbar}_{k-1,k} -{1\over 3}\eta^{i\jbar }
\Phi^l_{k,k-1}\Phi^{\bar l}_{k-1,k} \right)
e^{ i n\alpha (k-1)}~~,
\end{equation}
and singlets
\begin{equation}
O_n =\sum_{k=1}^3
\Phi^i_{k,k-1}\Phi^{\ibar}_{k-1,k}
e^{ i n\alpha (k-1)}~~,
\end{equation}
where $n$ assumes values $-1,0,1$.

Specifying the result (\ref{Zktotal})
of Appendix B to this $\IZ_3$ case, we find
that the $O(\lambda^2)$ source term in the octet beta function
$\beta_{{\bf 8},1}$ vanishes. Therefore, the operator
 $O_1^{\langle i{\ibar }\rangle}
    O_{-1}^{\langle j{\ibar }\rangle}$ is not generated along the
fixed line emanating from the origin of the coupling constant
space. However, (\ref{Zktotal}) does give an
$O(\lambda^2)$ source in the beta function for the
singlet operator $O_1 O_{-1}$.
To explain the physical meaning of
this, we recall that the standard orbifold projection method
yields $U(N)^k/U(1)$ theories which contain non-conformal $U(1)$
factors with abelian charges $e$ set equal to the $SU(N)$
charges $g_{\scriptscriptstyle YM}$. In \cite{Fuchs} this choice of parameters
was called the ``natural line.'' On this line the theory cannot be
conformal because $\beta_e$ is positive.
For example in the supersymmetric $\IZ_3$ case,
\begin{equation} \label{U1beta}
\beta_e = {3\over 16\pi^2 } e^3 N
\ .
\end{equation}
The contributions to the potential from the D-terms of the
$U(1)$ factors give
\begin{equation}
{e^2\over 2} \left (|\Phi^i_{1,2}|^2 -|\Phi^i_{1,3}|^2\right )^2+
{e^2\over 2} \left (|\Phi^i_{1,2}|^2 -|\Phi^i_{2,3}|^2 \right )^2+
{e^2\over 2} \left (|\Phi^i_{2,3}|^2 -|\Phi^i_{1,3}|^2\right )^2=
e^2 O_1 O_{-1}
\end{equation}
Thus, on the natural line the singlet double-trace operator
is automatically present in the tree-level action, with
coefficient $e^2$.
The flow of the abelian charge $e$ (\ref{U1beta}) then explains
why there must be a beta function generated for the singlet
double-trace operator.\footnote{The $U(1)$ factors manifest
themselves in an even more
dramatic fashion if the orbifold preserves ${\cal N}=2$ supersymmetry. In that
case, the chiral superfield in the twisted $U(1)$ ${\cal N}=2$ vector
multiplet appears in the tree level superpotential and leads to a
double-trace operator in a nontrivial representation of the
$R$-symmetry group already at tree level. In this case the flow of the
abelian charge should be responsible for the beta function of certain
nonsinglet operators. Clearly this additional subtlety is absent for
smaller amount of preserved supersymmetry. In fact, it is absent for
all theories where all scalar fields are bifundamentals, such
as the class of $SU(3)$ symmetric theories obtained by
(\ref{spinSU3}).}
The RG flow should take the theory from the
natural line $e=g_{\scriptscriptstyle YM}$ to the actual fixed line where $e=0$ and the
$U(1)$ factors decouple. By supersymmetry we then expect
(although have not checked in detail) that on the fixed line
no double trace operator
$O_{1} O_{-1}$ is generated.

In fact, the concern about the role of the $U(1)$ factors
in orbifold calculations for bifundamental scalars is general and
applies to both supersymmetric and non-supersymmetric examples.
 The simple method of
orbifold projection in field theory \cite{BJ} is very efficient
in producing results on the natural line, and we adopt it in our
paper. But to understand the
fixed line one needs more complicated calculations where the
$U(1)$ factors are decoupled and we have a $SU(N)^k$
theory. We note, however, that the $U(1)$'s affect only the
double-trace operators made out of $SU(3)$ singlets.
Thus, we can certainly take our results for $SU(3)$
adjoint beta functions $\beta_{{\bf 8},n}$ as applicable to the $SU(N)^k$
theory, and not just to the $U(N)^k/U(1)$ theory.

\subsection{$\IZ_6$}

\begin{figure}[ht]
\begin{center}
\epsfig{file=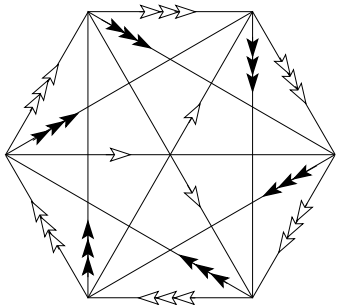,height=4cm}
\caption{ The quiver diagram for the non-supersymmetric freely acting
$\IZ_6$ orbifold. \label{Z6quiver}}
\end{center}
\end{figure}

The field content is quite similar to the one in the $\IZ_5$ theory,
so we will not describe it again. The quiver summarizing it is shown
in figure \ref{Z6quiver}.

It turns out that in this case not all octet operators are generated
at one-loop level:
$O_2^{\langle i{\jbar }\rangle}
O_{-2}^{\langle j{\ibar}\rangle}$ is not generated.
This can be traced to the fact that the orbifold action has
a $\IZ_3$ subgroup $\{1,g^2, g^4\}$ which
preserves the ${\cal N}=1$ supersymmetry.
Therefore, we may choose to deform the tree-level
action only with
\begin{eqnarray}
\delta_{\rm 2~trace}S&=&
 f_{{\bf 8},1}O_1^{\langle i{\jbar}\rangle}
    O_{-1}^{\langle j{\ibar }\rangle}
+\frac{1}{2}f_{{\bf 8},3} O_3^{\langle i{\jbar }\rangle}
    O_{-3}^{\langle j{\ibar }\rangle}
+f_{{\bf 1},1}O_1
    O_{-1}
+f_{{\bf 1},2} O_2
    O_{-2}
+\frac{1}{2}f_{{\bf 1},3} O_3
    O_{-3}
\end{eqnarray}

\begin{eqnarray}
\beta_{{\bf 8},1}&=&
\frac{1}{192\pi^2 }
\left[
(f_{{\bf 8},1}+12\lambda)^2 -80 {\lambda^2}\right]
\cr
\beta_{{\bf 8},3}&=&
\frac{1}{192\pi^2}
\left[
f_{{\bf 8}, 3}^2+64 {\lambda^2} \right]\cr
\beta_{{\bf 1},1}&=&
\frac{1}{64\pi^2}
\left[
(f_{{\bf 1}, 1}+6\lambda)^2- \frac{320}{9}\lambda^2 \right]\cr
\beta_{{\bf 1},2}&=&
\frac{1}{64\pi^2 }
\left[(f_{{\bf 1}, 2}+6\lambda)^2\right]\cr
\beta_{{\bf 1},3}&=&
\frac{1}{64\pi^2 }
\left[
(f_{{\bf 1}, 3}+4\lambda)^2+ \frac{112}{9}\lambda^2 \right]
\end{eqnarray}

Thus, we see that only the beta functions for the highest charge
operators do not posses a nontrivial zero at real values of the
double-trace coupling constant.

As we will see however in the next subsection, the fact that only a
small number of couplings have positive beta functions is nongeneric
within the class of orbifolds considered here; roughly speaking, about
one quarter of all double-trace operators have this unfortunate property.

\subsection{$\IZ_k$ with $SU(3)$ symmetry}

It is relatively easy to specialize the results obtained in
\S\ref{gen_orb} to the case of $\IZ_k$ with $SU(3)$ symmetry and we
spell this out in the Appendix \ref{ZkSU3}. 
The point worth emphasizing here is
that, similarly to the $\IZ_5$ and $\IZ_6$ examples,
the double-trace tree-level terms are:
\begin{eqnarray}
\delta S^{\rm tree}_{\rm 2~trace}&=&
\frac{1}{2}\sum_{n=1}^{k-1}\,f_{{\bf 8},n}
O_n^{\langle i\jbar \rangle} O_{-n}^{\langle j\ibar \rangle}
+
\frac{1}{2}\sum_{n=1}^{k-1}\,f_{{\bf 1},n} O_n O_{-n}
\end{eqnarray}
with the symmetry
\begin{eqnarray}
f_{{\bf 8},n} = f_{{\bf 8},k-n}~~~~~~~~ f_{{\bf 1},n} = f_{{\bf 1},k-n}
\end{eqnarray}
This choice of coefficients leads to unified expressions for the beta
functions for the couplings $f_{{\bf 1}, n}$ and $f_{{\bf 8}, n}$ for all values
of the charge. The calculations described in the Appendix \ref{ZkSU3}
leads to the following beta functions:
\begin{eqnarray}
\beta_{8, n}&=&\frac{1}{32\pi^2 k}
\left[(f_{8,n}+2\delta_{8,n})^2
-256 \lambda^2 \left(3+4\cos (n\alpha_k)\right)
\sin^4 \left ({\textstyle{\frac{1}{2}}}n\alpha_k\right )
\right]
\label{betaSU3}
\\
\beta_{1, n}&=&\frac{3}{32\pi^2 k}
\left[\left(f_{1,n}+\frac{2}{3}\delta_{1,n}\right)^2
-\frac{64}{9} \lambda^2 \left(1+2\cos (n\alpha_k)\right)
\left(4-2\cos (n\alpha_k)
+\cos (2n\alpha_k)\right)
\right]\nonumber
\end{eqnarray}

$\!\!\!\!$\mbox{
{
\epsfig{file=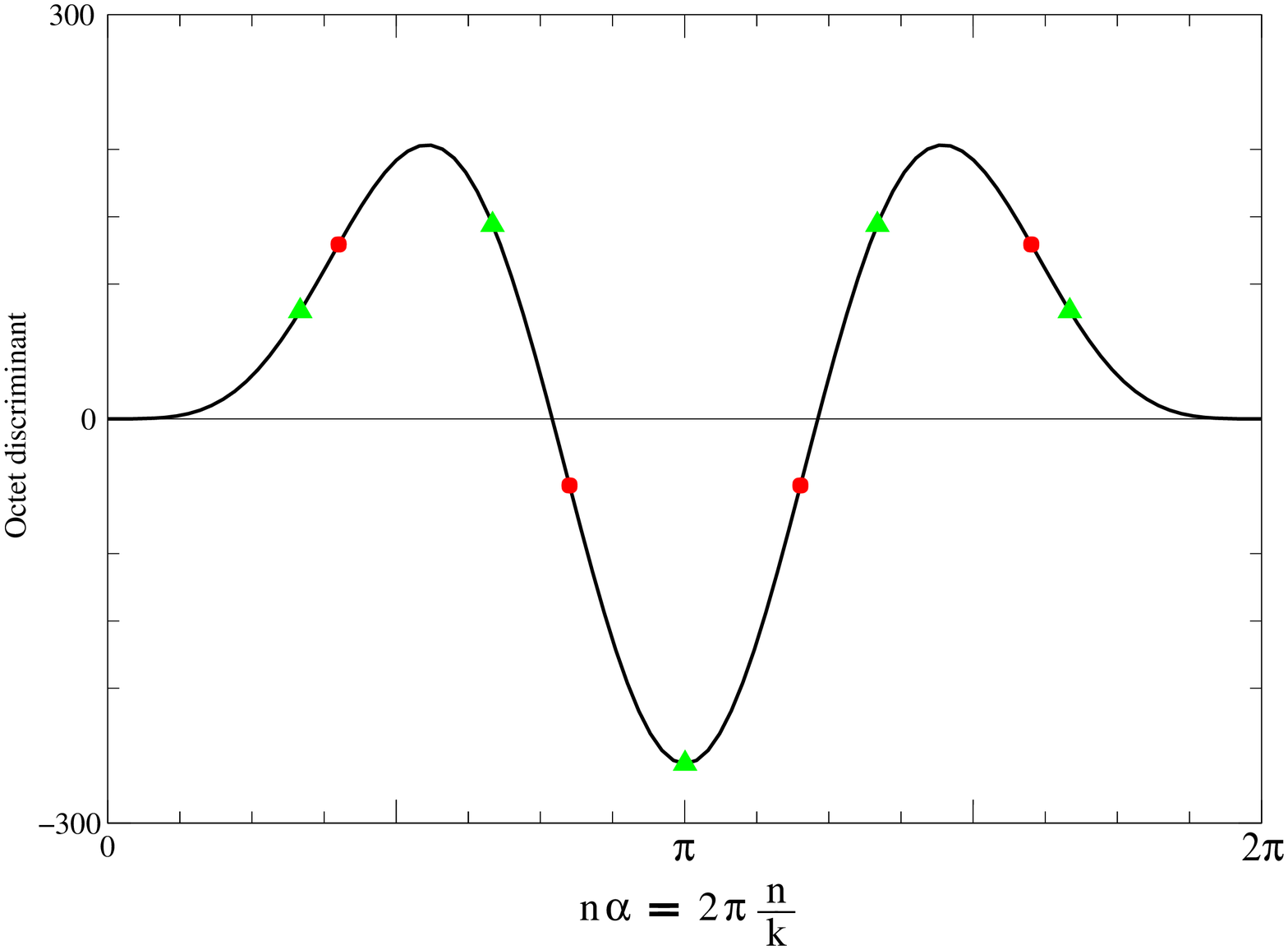,height=5.2truecm}}
}
~~~~~
\raisebox{2.46truecm}{\mbox{\begin{minipage}{7.3cm}
{
{
\epsfig{file=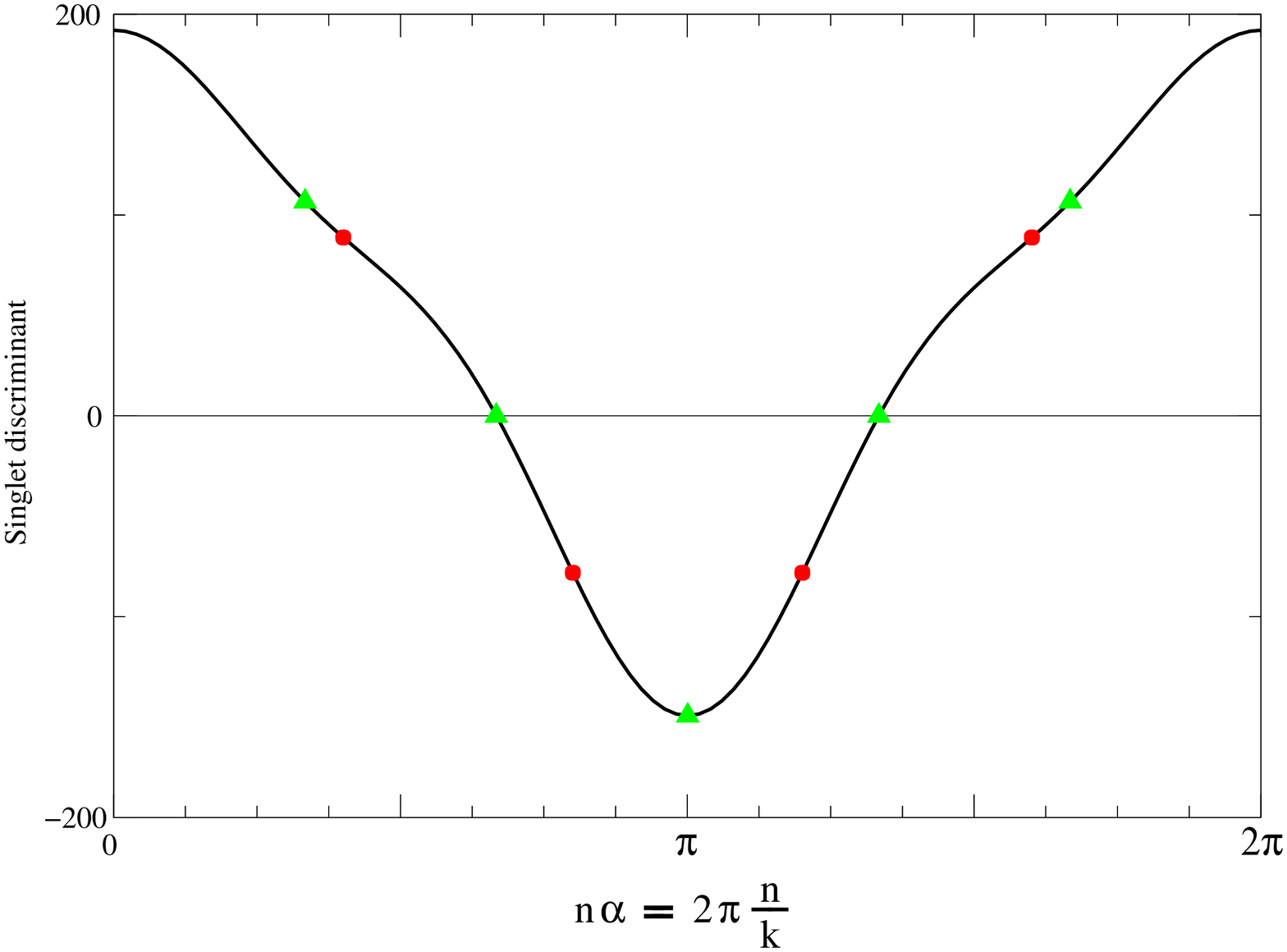,height=5.28truecm}}
}
\end{minipage}}}

The figures above represent plots of the discriminants of the
equations imposing the vanishing of the octet and singlet beta
functions, respectively. The circular dots and upright triangles
correspond to the $\IZ_5$ and $\IZ_6$ examples discussed in the
beginning of this section.

{}From (\ref{betaSU3}) we see that the charge $m$ operators appearing
in the $\IZ_{2m}$ theory are special in that, up to the factor of
$1/|\Gamma|$, the beta functions for their coefficients are equal.
This is the manifestation of the fact emphasized in the beginning
that, in a sense, this operator can be thought of as being inherited
from a $\IZ_2$-orbifold field theory analogous to the one discussed
in \S\ref{Z2example}.

Another important point is that all orbifold field theories of the
type discussed in this section posses at least one
deformation made out of $SU(3)$ adjoints which
spoils conformal symmetry. Indeed, the beta function of
all such operators with
\begin{eqnarray}
-1\le \cos \left(2\pi{{\frac{n}{k}}}\right) < -\frac{3}{4}
\end{eqnarray}
has a negative discriminant and hence no real solutions for the
coupling constant. Such operators are not affected by
the contributions of the twisted $U(1)$'s, which need to be removed
from the orbifold theory because they become free in the IR.
Therefore, even without performing the more
laborious calculations which separate out the $U(1)$'s, we conclude
that there is no fixed line passing
through the origin in the non-supersymmetric theories with
the $SU(3)$ symmetry.


\section{Discussion}

In this paper we searched for perturbative fixed lines in
$d=4$ non-supersymmetric large $N$ orbifold gauge theories.
This required a careful calculation
of the one-loop beta-functions for double-trace operators made
of scalar fields. In the examples we considered,
both freely acting and not,
we found that
there are no fixed lines passing through
the origin in the coupling constant space
$(f^i,\lambda)$. We hope to return to perturbative
analysis of more general gauge theories in the future.

Let us consider how our one-loop results may be modified by
higher order corrections.
The two-loop correction to $\beta_f$ has
the general structure
\begin{equation} \label{genbetanew}
\delta \beta_f=  w f^3 + v_1 \lambda f^2 + 2 \gamma_1 f \lambda^2
+ a_1 \lambda^3
\ .
\end{equation}
For example, the third term arises through a two-loop correction to the
anomalous dimension of $O$, $\gamma_1\lambda^2$.
For $f$ of order $\lambda$ these terms are of order $\lambda^3$
and are suppressed at weak coupling compared to the one-loop
contributions.
Thus, the two-loop correction is small near the origin
of the coupling constant space. If a real solution
$f=a \lambda$ exists at one-loop order then it should be possible to
correct it, $f= a\lambda + b\lambda^2$, so that the two-loop
solution exists to order $\lambda^3$. Iterating this procedure
order by order, we would conclude that if a real one-loop
solution exists then there is indeed a fixed line passing through
the origin of the coupling constant space.

 In cases where one-loop
beta function equation has no real solutions, we cannot rule out
a possibility that there is a fixed line passing away from the origin,
although it is difficult to study perturbatively.
Another possibility is that there is a line of fixed points which
terminates at some critical value
$\lambda_c$ without reaching the weak coupling region.
Indeed, for freely acting orbifolds, such as the $SU(3)$ symmetric family
we considered,
there is evidence from AdS/CFT for a fixed line
at very large values of $\lambda$ where there are no tachyons in the twisted
sector.\footnote{
Even for non-freely acting orbifolds
it seems possible
that there is a fixed line passing away from the origin or terminating
at intermediate coupling.
In this case, however,
the AdS/CFT correspondence indicates that the theory
is unstable for large $\lambda$. So, at best, the large $N$ theory
stays conformal only for a certain range of $\lambda$.}
We will comment on a possible mechanism for a phase transition at $\lambda_c$
later in this section. In any case,
our perturbative calculation shows that this fixed line
cannot pass through the origin of the coupling constant space, 
but it would be interesting
to examine more general freely acting orbifolds.

For instance, certain
product group orbifolds, such as $\IZ_m\times \IZ_n$ orbifolds,
appear promising in this
direction. In our discussion of the $\IZ_n$ orbifolds with $SU(3)$
symmetry we saw that, in some cases, the beta functions without fixed
points are associated with the operators charged under some
non-freely acting subgroup of the full orbifold group. Thus, a step
toward finding a non-supersymmetric conformal orbifold in the large $N$
limit could be ensuring that such operators are not generated. For a
product group orbifold we may choose that each of the individual
factors acts in a supersymmetric way, but preserving different
subalgebras of the ${\cal N}=4$ supersymmetry algebra.
One example is the freely acting $\IZ_3\times \IZ_3$ orbifold where the
first $\IZ_3$ is generated by
$r= (\omega_3,\omega_3,\omega_3,1)$ and the second by
$r'= (1, \omega_3,\omega_3,\omega_3)$. Thus, each
$\IZ_3$ preserves ${\cal N}=1$ supersymmetry by itself, but
the combined orbifold breaks all supersymmetry.
In this class of orbifolds, some double-trace operators are
not generated due to the supersymmetry of the individual factors.
It would be very interesting to examine all double-trace
beta functions,
after eliminating by hand the twisted $U(1)$ factors.


Finally, we suggest a relation of our calculations to
closed string tachyon condensation. In freely acting orbifolds of
$AdS_5\times S^5$, all twisted sector tachyons are lifted
at large $\lambda$ because the twisted strings
are highly stretched. Schematically, the effective
potential for such a charged field in $AdS_5$ has the form
\begin{equation} \label{stringpot}
V (T)= {1\over 2} m^2 (\lambda) T^* T + {c_4\over 4} (T^* T)^2 + \ldots
\ ,
\end{equation}
where at strong coupling $m^2 (\lambda) \sim \lambda/\alpha'$.
In \cite{Dine,Suyama} a similar effective potential was studied
in a simpler case of the Rohm compactification,
where $T$ corresponds to a string winding around a circle.
In that case $c_4$ was found to be positive;
we will assume that it is positive also in (\ref{stringpot}).
Since the twisted sector strings have negative zero-point energy,
$m^2(\lambda)$ is expected to become negative for $\lambda < \lambda_c$,
where $\lambda_c$ is of order $1$.\footnote{
The position of the phase transition is affected by
the higher-derivative terms that may be present in the action,
for example $T^* T R_{abcd}^2$.} This may cause a transition to
a phase where $T$ has an expectation value 
(however, the
local maximum of the potential at $T=0$
is stable if $m^2$ is not too negative,
due to the well-known Breitenlohner-Freedman bound in AdS space
\cite{BF}).
What would be a manifestation of the tachyon condensation
in the dual gauge theory? The logarithmic running 
of double-trace
operators $O_n O_{-n}$ in the gauge theory
is expected to lead to development of expectation values
for twisted operators $O_n$ through the Coleman-Weinberg mechanism.
 This breaking of the quantum symmetry $\Gamma$
has been proposed \cite{Adams}
as the gauge theory dual of tachyon condensation.
We believe that this mechanism is applicable to both
non-freely acting and freely acting orbifolds.
Through an explicit gauge theory calculation we find that
at weak coupling all possible such trace-squared operators
typically get induced.
But the arguments based on the AdS/CFT correspondence suggest that,
as $\lambda$ is increased, the gauge theory should make a transition
to a phase with restored symmetry, i.e. with no 
running of double-trace operators.
This can happen if for sufficiently large `t Hooft
coupling $\lambda$ all the double trace beta
functions acquire real zeros.

If this scenario holds, then for 
sufficiently large $\lambda$ the perturbative
expansion probably breaks down, and the theory enters a different phase
which is dual to string theory on $AdS_5\times S^5/|\Gamma|$ with
radius larger than critical.
In large-$N$ theories this is a common situation since
summation of planar graphs typically has a finite radius of
convergence.\footnote{There are many known examples
in large-$N$ matrix models: for example, the Gross-Witten transition
\cite{Gross},
whose relation to non-supersymmetric string theory was proposed in
\cite{KMS}.} It is tempting to speculate that this kind of
large-$N$ phase transition in non-supersymmetric 
freely acting orbifold
gauge theories corresponds to symmetry restoration.

As we emphasized throughout the paper,
 there are two possibilities for the behavior of the theory
at very weak coupling. 
The first possibility is that some of the $a^i_\pm$ are complex.
Then for physical (real) values of double-trace couplings, they
do not possess
fixed points in the perturbative regime and flow away from it,
leading to runaway behavior in the large $N$ limit.
The examples we have considered
so far exhibit only this effect, which is a rather uncontrollable
tachyon condensation. 
The second possibility is that all $a^i_\pm$ are real.
Then all double-trace couplings flow to zeros
of their beta functions, and we have a fixed line passing through the origin
of the coupling constant space. In such a theory there 
does not seem to be a phase transition that could correspond to tachyon 
condensation. 
It remains to be seen whether there exist examples of such theories.

\section*{Acknowledgments}
We are grateful to S. Gubser, A. Polyakov and
E. Witten for very useful discussions and to H.~Schnitzer for
pointing out reference \cite{SCHN}.
The research of A.~D.  is
supported in part by grant RFBR 04-02-16538
and the National Science Foundation Grant No.~PHY-0243680.
The research of I.~R.~K. is
supported in part by the National Science Foundation Grant
No.~PHY-0243680. 
The research of R.~R. is supported in part by funds provided by the
U.S.D.O.E.
under co-operative research agreement
DE-FC02-91ER40671.
Any opinions, findings, and conclusions or recommendations expressed in
this material are those of the authors and do not necessarily reflect
the views of the National Science Foundation.

\newpage


\appendix


\section{Orbifold traces of Dirac matrices}

As we have seen in \S\ref{gen_orb}, the contribution of the
fermions to the effective action depends on the tensor
\begin{eqnarray}
\Tr[\gamma^I\gamma^J\gamma^K\gamma^Lr_g]~~,
\end{eqnarray}
where $\gamma$ are six-dimensional chiral (i.e. $4\times 4$)
Dirac matrices and $r_g$ is the realization of the orbifold group
element $g$ on the fundamental representation of $SU(4)$.

This trace can be computed for a general $r_g$-matrix. The idea is to
use the explicit form of the Dirac matrices, which can be inferred from
the fact that they realize the map between the ${\bf 6}$ of $SO(6)$
and $SU(4)$. This map naturally splits the $SO(6)$ vector indices
into complex three-dimensional indices: $I=(i,\,{\ibar })$. Also, this
map singles out the fourth component of the fundamental representation
of $SU(4)$ and splits the four-dimensional index as $\alpha=(i,4)$.

The nonvanishing traces are:
\begin{eqnarray}
\frac{1}{4}Tr[\gamma^{ \ibar }\gamma^{ \jbar }\gamma^{ k }\gamma^{ l }r_g]
&=&
   \epsilon^{klx}r_x{}^y\epsilon_{yji}\cr
\frac{1}{4}Tr[\gamma^{ i }\gamma^{ \jbar }\gamma^{ k }\gamma^{\bar l}r_g]
&=&
   (\delta^i_j\delta^k_lr_4{}^4+\delta^k_j r_l{}^i)\cr
\frac{1}{4}Tr[\gamma^{ i }\gamma^{ \jbar }\gamma^{\bar k}\gamma^{ l }r_g]
&=&
   (r_j{}^i\delta_k^l-r_k{}^i\delta_j^l)\cr
\frac{1}{4}Tr[\gamma^{\ibar }\gamma^{ j }\gamma^{ k }\gamma^{\bar l }r_g]
&=&
   (r_l{}^k\delta_i^j-r_l{}^j\delta_i^k)\cr
\frac{1}{4}Tr[\gamma^{\ibar}\gamma^{ j }\gamma^{\bar k}\gamma^{l}r_g]
&=&
   (r_x{}^x\delta_i^j\delta_k^l+\delta_i^lr_k{}^j
    -\delta_k^lr_i{}^j-\delta_i^jr_k{}^l) \cr
\frac{1}{4}Tr[\gamma^{ i }\gamma^{ j }\gamma^{\bar k}\gamma^{\bar l}r_g]
&=&
   r_4{}^4(\delta_l^i\delta_k^j-\delta_k^i\delta_l^j)\cr
\frac{1}{4}Tr[\gamma^{ i }\gamma^{ \jbar }\gamma^{ k }\gamma^{ l }r_g]
&=&
   \delta_j^i\epsilon^{lkx}r_x{}^4\cr
\frac{1}{4}Tr[\gamma^{\ibar}\gamma^{ j }\gamma^{\bar k}\gamma^{\bar l}r_g]
&=&
   (r_4{}^y\delta_i^j-r_4{}^j\delta_i^y)\epsilon_{ylk}\cr
\frac{1}{4}Tr[\gamma^{ i }\gamma^{ j }\gamma^{\bar k}\gamma^{ l }r_g]
&=&
   (\delta_x^i\delta_k^j-\delta_x^j\delta_k^i)\epsilon^{lxy}r_y{}^4\cr
\frac{1}{4}Tr[\gamma^{\ibar }\gamma^{\jbar }\gamma^{k}\gamma^{\bar l}r_g]
&=&
   \delta^k_l r_4{}^y\epsilon_{yij}\cr
\frac{1}{4}Tr[\gamma^{ i }\gamma^{ j }\gamma^{ k }\gamma^{\bar l}r_g]
&=&
   \epsilon^{ikj}r_l{}^4\cr
\frac{1}{4}Tr[\gamma^{ i }\gamma^{\jbar}\gamma^{\bar k}\gamma^{\bar l}r_g]
&=&
   r_4{}^i\epsilon_{jkl}
\end{eqnarray}
For abelian orbifolds one may diagonalize simultaneously
all the group generators and thus one may pick $r_g$ to be diagonal.
Then, the expressions above simplify considerably, in part due to the
vanishing of the last six lines.

\section{Renormalization of a single-trace operator\label{gen_anom_dim}}

The action of the 1-loop dilatation operator on long twisted operators
in the SU(2) sector was discussed in \cite{IDEG}. 
However, this discussion does not directly apply to operators of
dimension 2, because in this case there exist additional planar
diagrams. Furthermore, we are interested in more general operators
than those present in the $SU(2)$ sector.

The most general single-trace dimension 2 twisted operator is
\be
\Tr(g\phi^I \phi^J )~~.
\ee
The action of the dilatation operator is easy to find:
\begin{eqnarray}
&&\Delta(\Tr(g\phi^I \phi^J ))={\lambda\over
16\pi^2 \,|\Gamma|}\Big[\Tr(g\phi^I \phi^J )- \cr &&
-{1\over|\Gamma|}\sum_{{\tilde g}_1,{\tilde g}_2\in \Gamma}
(R_{{\tilde g}_1})^{I'I} (R_{{\tilde g}_2})^{J'J}
\left(\delta^{KJ'}\delta^{LI'}-{1\over
2}\delta^{KI'}\delta^{LJ'}-{1\over
2}\delta^{I'J'}\delta^{KL}\right)
\times  \cr
&&~~~~~~~~~~~~~~~~~~~~~~~~~~~~~~~~
\times
\left(\Tr(g {\tilde g}_1 \phi^L\phi^K {\tilde g}_1^\dagger)
\delta_{{\tilde g}_2=g{\tilde g}_1}+
\Tr(g{\tilde g}_2\phi^K\phi^L {\tilde g}_2^\dagger)
\delta_{{\tilde g}_1={\tilde g}_2}\right) \Big]
\label{diags}
\end{eqnarray}

This expression can be organized in various ways, each form
emphasizing different aspects of $\Delta$. For example,
\be
\Delta(\Tr(g\phi^I\phi^J))=
{\lambda\over 8\pi^2|\Gamma|}
\left[\Tr(g[\phi^I,g]g^\dagger\phi^J)+\Tr(g[\phi^I,\phi^J])+{1\over
2}(\delta^{IJ}+R_g{}^J{}^I)\sum_K \Tr(g\phi^K\phi^K)\right]
\ee
implies that traceless $\Gamma$-invariant bilinears have zero anomalous
dimension at one loop.

Alternatively, (\ref{diags}) can be cast in a form very similar to the
dilatation operator in the parent theory
\begin{eqnarray}
\Delta\left(\Tr(g\phi^I\phi^J)\right)&=&
{\lambda\over 16\pi^2|\Gamma|} \bigg [
4\Tr(g\phi^I\phi^J)+\left(\delta^{IJ}+R_g^{JI}\right)\sum_K
\Tr(g\phi^K\phi^K)\cr
&-&2\left(\Tr(g\phi^I\phi^K)\left(R_g\right){}^{KJ}
+\left(R_g^{-1}\right){}^{IK}\Tr(g\phi^K\phi^J)\right) \bigg ]~~,
\label{eq:gen_anom_dim}
\end{eqnarray}
which makes it easy to compute the anomalous dimensions in cases in
which $R_g$ is diagonal.

\section{$\IZ_k$ with $SU(3)$ symmetry \label{ZkSU3}}

\subsection{Anomalous dimensions}

It is easy to see from (\ref{eq:gen_anom_dim})
that the ``off-diagonal operators''
-- $\Tr(g^n \,\phi^i\phi^{\jbar})$ with $i\ne j$ -- have definite
anomalous dimensions:
\begin{eqnarray}
\Delta \Tr(g^n \,\phi^i\phi^{\jbar})\Big|_{i\ne j}
\!\!
=\frac{\lambda}{16\pi^2 k}\left[
\,8\sin^2 (n\alpha_k)\,
\Tr(g^n \,\phi^i\phi^{\jbar})\Big|_{i\ne j}\right]\ .
\end{eqnarray}
The diagonal operators however -- $\Tr(g^n \phi^i\phi^{\ibar})$ --
mix under RG flow. Their mixing matrix is
\begin{equation}
\Delta{\cal O}_n=\frac{\lambda}{16\pi^2 k}
\begin{pmatrix}
4\sin^2 (n\alpha_k) +4 & 4\cos^2 (n \alpha_k)
& 4\cos^2 (n \alpha_k) \cr
4\cos^2 (n \alpha_k) & 4\sin^2 (n\alpha_k) +4
& 4\cos^2 (n\alpha_k) \cr
4\cos^2 (n \alpha_k) & 4\cos^2 (n \alpha_k)
& 4\sin^2 (n\alpha_k) +4
\end{pmatrix}{\cal O}_n
\end{equation}
where
\begin{equation}
{\cal O}_n=\begin{pmatrix}
\Tr(g^n\phi^1\phi^{\bar 1})\cr
\Tr(g^n\phi^2\phi^{\bar 2})\cr
\Tr(g^n\phi^3\phi^{\bar 3})
\end{pmatrix}
\end{equation}
Its diagonal form is:
\begin{eqnarray}
\Delta{\widetilde {\cal O}}_n=\frac{\lambda}{16\pi^2 k}
\begin{pmatrix}
2\left(5+\cos (2n\alpha_k)\right)
& 0
& 0  \cr
  0
& 8\sin^2 (n\alpha_k)
& 0 \cr
  0
& 0
& 8\sin^2 (n\alpha_k)
\end{pmatrix}{\widetilde {\cal O}}_n
\end{eqnarray}
For $n=0$ we recover the well-known value of the 1-loop
anomalous dimension of the Konishi operator.
Collecting everything we find the operators and dimensions
quoted in (\ref{anomdimSU3}).

\subsection{Effective action}

The bosonic and ghost loop contribution is
\begin{eqnarray}
\delta S^{\rm 1~loop|2~tr}_{\rm Bose,\,ghost}&=&-\frac{\Div}{2\,k}
\sum_{n=0}^{k-1}
\Big\{\hphantom{-}
8\big(\cos^2(n\alpha_k)\left(3\cos(2n\alpha_k)+1\right)+
2\sin^2(2n\alpha_k)\big)
O_nO_{-n}  \cr
&&~~~~~~~~~~~~~~
+8\big(2+\cos(2 n \alpha_k )+\cos(4 n \alpha_k )\big)
O^{i\jbar }_nO^{j\ibar }_{-n}
~~~~\Big\}
\label{Z5bgh}
\end{eqnarray}

The fermionic contribution is
\begin{eqnarray}
\delta S^{\rm 1~loop|2~tr}_{\rm Fermi}&=&\frac{\Div}{2\,k}
\sum_{n=0}^{k-1} \,\Big\{\hphantom{-}
8\big(4\cos^3(n\alpha_k) + \cos(n\alpha_k) - \cos(3n\alpha_k)\big)
O_n O_{-n} \cr
&&~~~~~~~~~~~~~
+8\big(4\cos^3(n\alpha_k) - \cos(n\alpha_k) + \cos(3n\alpha_k)\big)
O_n^{i{\jbar}}O_{-n}^{j\ibar }
~~~~\Big\}
\label{Z5f}
\end{eqnarray}

Adding the equations (\ref{Z5f}) and (\ref{Z5bgh}) and expressing the
result in terms of operators with definite 1-loop
scaling dimension leads to:
\begin{eqnarray} \label{Zktotal}
\delta S^{\rm 1~loop|2~tr}_{\rm \scriptscriptstyle Bose,\,ghost,Fermi}
&=&-\frac{\Div}{2\,k}
\sum_{n=1}^{k-1}
\Bigg[
\hphantom{+}
\frac{256}{3}\sin^8 \left ( {\textstyle{\frac{1}{2}}}n\alpha_k\right )
\, O_nO_{-n}
\cr
&&~~~~~~~~~~~~~~~~~~~~
+64 (1+2\cos (n\alpha_k))^2
\sin^4 \left ({\textstyle{\frac{1}{2}}}n\alpha_k\, \right )
O^{\langle i\jbar \rangle}_n O^{\langle j\ibar\rangle }_{-n}
~~~\Bigg]
\end{eqnarray}

We add the following double-trace terms to the
tree-level action:
\begin{eqnarray}
\delta S^{\rm tree}_{\rm 2~trace}&=&
\frac{1}{2}\sum_{n=1}^{k-1}\,f_{{\bf 8},n}
O_n^{\langle i\jbar \rangle} O_{-n}^{\langle j\ibar \rangle}
+
\frac{1}{2}\sum_{n=1}^{k-1}\,f_{{\bf 1},n} O_n O_{-n}
\end{eqnarray}
with the symmetry
\begin{eqnarray}
f_{{\bf 8},n} = f_{{\bf 8},k-n}~~~~~~~~
f_{{\bf 1},n} = f_{{\bf 1}, k-n}
\end{eqnarray}
For each $SU(3)$ structure there are $[k/2]$ independent couplings
and hence $[k/2]$ beta functions.

The contribution to the effective action:
\begin{eqnarray}
\delta
S^{\rm 1~loop|2~tr}_{\rm 2~trace}
&=&-\frac{\Div}{2\,k}\Bigg[\hphantom{+}
\sum_{n=1}^{k-1}\,\frac{f^2_{{\bf 8},n}}{4}
O_n^{\langle i\jbar \rangle} O_{-n}^{\langle j\ibar \rangle}
+
\sum_{n=1}^{k-1}\,\frac{3f^2_{{\bf 1},n}}{4} O_n O_{-n}
\cr
&&~~~~~~~~
+
\sum_{n=1}^{k-1}\,\delta_{{\bf 8},n}\,f_{{\bf 8},n}
O_n^{\langle i\jbar \rangle} O_{-n}^{\langle j\ibar \rangle}
+
\sum_{n=1}^{k-1}\,\delta_{{\bf 1},n}\,f_{{\bf 1},n} O_n O_{-n}\Bigg]
\end{eqnarray}

\begin{eqnarray}
\gamma_{{\bf r},n} = \frac{1}{16\pi^2\,k}\,\delta_{{\bf r},n}
\end{eqnarray}
are the anomalous dimensions of the operators with representation and
charge $({\bf r}, n)$.

The beta functions are then given by:
\begin{eqnarray}
\beta_{8, n}&=&\frac{1}{16\pi^2\,k}
\left[128(1+2\cos\left(n\alpha_k\right))^2
\sin^4\!\left({\textstyle{\frac{1}{2}}}n\alpha_k\right)\lambda^2
+\frac{f_{{\bf 8},n}^2}{2}
+2\delta_{{\bf 8},n}f_{{\bf 8},n}
\right]
\cr
\beta_{1, n}&=&\frac{1}{16\pi^2\,k}
\left[
\frac{512}{3}\sin^8\!\left({\textstyle{\frac{1}{2}}}n\alpha_k\right)
\lambda^2
+\frac{3f_{{\bf 1},n}^2}{2}
+2\delta_{{\bf 1},n}f_{{\bf 1},n}
\right]
\end{eqnarray}
It is then a simple exercise to use  the anomalous dimensions
(\ref{anomdimSU3}) and obtain the beta functions
(\ref{betaSU3}).

\section{$\IZ_k$ with fixed points and $SO(4)$ global
symmetry \label{SO4}}

In this appendix we summarize the effective action of the orbifold field
theories with $SO(4)$ global symmetry. They are \cite{ADPS}
the gauge theory
realizations of the well-known orbifolds $\IC^2\times \IC/\IZ_k$ with
$D3$ branes placed at the tip of the
cone (for a recent review see \cite{HEMT}).

\subsection{Spectrum}

As in the previous case, we choose the standard representation of
$\IZ_k$ on the gauge degrees of freedom. Furthermore, we will choose
the following action on the fundamental representation of
$SU(4)$ and the vector representation of $SO(6)$:
\begin{eqnarray}
r={\rm diag}(\omega_k,\omega_k,\omega_k^{-1},\omega_k^{-1})
\end{eqnarray}
\begin{eqnarray}
R={\rm diag}(1, 1, \omega_k^2,1, 1,\omega_k^{-2})
\end{eqnarray}
where the two entries different from unity are the weights of
$\phi^3$ and $\phi^{\bar 3}$, respectively.

We will use real indices for the uncharged fields and complex indices
for $\phi^3$ and $\phi^{\bar 3}$. Furthermore, the metric on this
space of fields is:
\begin{eqnarray}
\eta={\scriptsize{
\begin{pmatrix}
1& 0& 0& 0& 0& 0\cr
0& 1& 0& 0& 0& 0\cr
0& 0& 1& 0& 0& 0\cr
0& 0& 0& 1& 0& 0\cr
0& 0& 0& 0& 0& 1\cr
0& 0& 0& 0& 1& 0
\end{pmatrix}
}}
\end{eqnarray}

The invariant fields allow the construction of the following
operators:
\begin{eqnarray}
O_n^{\langle \mu\nu\rangle}=
\Tr(g^n\phi^\mu\phi^\nu)-\frac{1}{4}\delta^{\mu\nu}O_n\ ,
~~~~~~~
O_n=\sum_{a=1}^4\Tr(g^n\phi^\kappa\phi^\kappa)\ ,
~~~~~~~
A_n=\omega_k^n\Tr(g^n\phi^3\phi^{\bar 3})\ .
\end{eqnarray}
The reason for the apparently strange phase in $A_n$ is that
with this normalization the entries of the anomalous dimension matrix
are real.

\subsection{Effective action}

The bosonic, ghost and fermionic contribution to the effective action is:
\begin{eqnarray}
\label{effso4bgf}
\delta
S^{\rm \scriptscriptstyle{1~loop|2~tr}}_{\rm \scriptscriptstyle{b,gh,f}}
&=&-\frac{\Div}{2\,k}\sum_{n=1}^{k-1}
\Bigg[\hphantom{+}
16\sin^4 ({\textstyle{\frac{1}{2}}}n\alpha_k)
\left(2O_n^{\langle \mu\nu\rangle}
O_{-n}^{\langle \mu\nu\rangle}+\frac{3}{2}O_nO_{-n}\right)\\
&&~~~~~~~~~~~~~~
+32\left(7+4\cos(n\alpha_k)+\cos(2n\alpha_k)\right)
\sin^4\left({\textstyle{\frac{1}{2}}}n\alpha_k\right)
A_n A_{-n}\cr
&&~~~~~~~~~~~~~~
-32(2+\cos(n\alpha_k))
\sin^4\left({\textstyle{\frac{1}{2}}}n\alpha_k\right)
\left(O_n A_{-n} +O_{-n} A_{n}\right)
~~~~~
\Bigg]\nonumber
\end{eqnarray}

\begin{eqnarray}
\delta_{\rm {\scriptscriptstyle 2\,tr}}S 
=
\frac{1}{2}\sum_{n=1}^{k-1}\,f_{{\bf 9},n}
O_n^{\langle \mu\nu \rangle} O_{-n}^{\langle \mu\nu \rangle}
+
\frac{1}{2}\sum_{n=1}^{k-1}\,f_{{\bf 1},n} O_n O_{-n}
+
\frac{1}{2}\sum_{n=1}^{k-1}\,f_{(3),n}
A_n A_{-n}
+
\sum_{n=1}^{k-1}\,
\,f_n\,O_n
A_{-n}
\end{eqnarray}
with constraints similar to those discussed before:
\begin{eqnarray}
f_{{\bf 9},n}=f_{{\bf 9},k-n}~~~~f_{{\bf 1},n}=f_{{\bf 1},k-n}~~~~
f_{(3),n}=f_{(3),k-n}~~~~~~~~f_{n}=f_{k-n}
\end{eqnarray}
where the last constraint ensures that the deformation of the tree
level action is real.

\begin{eqnarray}
\delta S^{\rm 1~loop|2~tr}_{\rm 2~trace}
&=&-
\frac{\Div}{2k}\Bigg\{\hphantom{+}
\sum_{n=1}^{k-1}\,\frac{f^2_{{\bf 9},n}}{2}
O_n^{\langle \mu\nu \rangle} O_{-n}^{\langle \mu\nu \rangle}
+
\sum_{n=1}^{k-1}\,\left[2{f^2_{{\bf 1},n}}+\frac{1}{4}f_nf_{k-n}\right]
O_n O_{-n} \cr
&&~~~~~~~~~~
+
\sum_{n=1}^{k-1}\,
\left[\frac{f^2_{(3),n}}{4}+2f_nf_{k-n}\right]
A_n A_{-n}
+
2\left[2f_{{\bf 1},n}f_n+\frac{1}{4}f_{(3),n}f_n\right]\,O_n
A_{-n}
\cr
&+&
\sum_{n=1}^{k-1}\,
\left[f_{{\bf 1},n} \left[\delta_n^{OO}O_n
+\delta_n^{OA}
A_{n}\right] O_{-n}
+
f_{(3),n} \left[\delta_n^{AO}O_n
+\delta_n^{AA}
A_n\right]
A_{-n}\right] \cr
&+&
\sum_{n=1}^{k-1}\,
\,f_n\,\left[\left[\delta_n^{OO}O_n
+\delta_n^{OA}
A_n\right]
A_{-n}+O_n
\left[\delta_{-n}^{AO}O_{-n} +\delta_{-n}^{AA}
A_{-n}\right]
\right]~~~
\Bigg\}
\label{2trso4}
\end{eqnarray}
where the anomalous dimension matrix is:
\begin{eqnarray}
\Delta
\begin{pmatrix}
O_n\cr A_n
\end{pmatrix}
=\frac{1}{16\pi^2 k}\delta
\begin{pmatrix}
O_n\cr A_n
\end{pmatrix}
=
\frac{\lambda}{16\pi^2k}
\begin{pmatrix}
8&16 \cos(n\alpha_k)\cr
2\cos(n\alpha_k) & 4\left(1+\sin^2(n\alpha_4)\right)
\end{pmatrix}
\begin{pmatrix}
O_n\cr A_n
\end{pmatrix}
\end{eqnarray}

To extract the beta functions we add (\ref{effso4bgf}) and
(\ref{2trso4}) while properly keeping track of the
charges or operators to eliminate the doubling introduced to
make the expressions uniform.
It is clear that $\beta_{{\bf 9},n}$ do not have any real zeros, since
the corresponding operators have no one-loop anomalous dimensions.
This agrees with our general findings for non-freely acting orbifolds
in section 4.3.

\end{document}